\documentclass[journal]{IEEEtran}

\usepackage{dcolumn}
\usepackage{bm}
\usepackage{textcomp}
\usepackage{chngcntr}
\usepackage{amsmath,amssymb,amsfonts}
\usepackage[percent]{overpic}
\usepackage{enumerate}
\usepackage{mathrsfs}
\usepackage{balance}
\usepackage{mathtools}
\usepackage{xr}
\usepackage{afterpage}
\usepackage{cases}
\usepackage{scalerel}
\usepackage{tabularx}
\usepackage{amsfonts}
\usepackage{xcolor}
\usepackage{cite}
\newcommand{\pvec}[1]{\vec{#1}\mkern2mu\vphantom{#1}}
\usepackage[mathscr]{euscript}
\usepackage{subfigure}
\usepackage{gensymb}
\usepackage{soul}

\begin{document}

\title{Extreme Beam-forming with Impedance Metasurfaces Featuring Embedded Sources
and Auxiliary Surface Wave Optimization}

\author{Gengyu Xu, \textit{Student Member, IEEE}, Vasileios G.~Ataloglou, \textit{Student Member, IEEE},\\ Sean V. Hum, \textit{Senior Member, IEEE} and George V. Eleftheriades, \textit{Fellow, IEEE}}

\maketitle

\begin{abstract}
We present the end-to-end design of compact passive and lossless metasurface (MTS) antennas with integrated feeds. The complete low-profile system consists of a single-layered reactive impedance MTS printed on top of a grounded dielectric substrate, and is fed by sources which are embedded inside the substrate. An accurate and efficient volume-surface integral equation-based model of the device is developed, and used as the basis for the rapid optimization of its performance. The optimized designs leverage tailored auxiliary surface waves supported by the impedance MTS to distribute the localized source power across their apertures. This facilitates the realization of extreme field transformations such as wide-angle beam-forming or shared-aperture beam-forming, with nearly 100\% aperture efficiencies. The procedure also allows for arbitrary beam-shaping with complete main beam and side lobe control. 
%We present a simple method to translate theoretically synthesized surfaces into realistic devices that can be easily fabricated using standard printed circuit board technology. 
We also derive several feasibility-related constraints, which can significantly enhance the power efficiency as well as the bandwidth of the MTS antennas when they are implemented in practice. Full-wave numerical simulations confirm the effectiveness of the presented approach, as well as the extreme field transformation capabilities of the synthesized designs.
\end{abstract}


%%%%%%%%%%%%%%%%%%%%%%%%%%%%%%%%%%%%%%%%%%%%%%%%%%%%%%
%%%%%%%%%%%%%%%%%%%%%%%%%%%%%%%%%%%%%%%%%%%%%%%%%%%%%%
\section{Introduction}
The research and development of antennas with extraordinary beam-forming capabilities have seen a resurgence owing to the advent of electromagnetic metasurfaces (MTSs)~\cite{roadmap:JOP2019}, which can be viewed as the two-dimensional analogue of metamaterials. They consist of dense arrays of scatterers (called meta-atoms) with judiciously designed electromagnetic polarizabilities. Due to their highly subwavelength thicknesses, MTSs can be theoretically modeled as infinitesimally thin sheets of effective electric and/or magnetic currents imposing abrupt discontinuities in the magnetic and/or electric fields. Full control over both fields can be achieved with the use of Huygens' metasurfaces (HMSs), which are electrically and magnetically polarizable~\cite{Selvanayagam:OptExp2013,Pfeiffer:PRL2013}. Even more types of field-transforming functionalities can be obtained by introducing cross-coupling between the electric and magnetic responses, thereby realizing so-called bianisotropic Huygens' metasurfaces (BMSs)~\cite{Pfeiffer:PRApplied2014,Achouri:TAP2015,Epstein:TAP2016,Asadchy:Nanophotonics2018}.

The strong and efficient interaction between radio frequency electromagnetic waves with metals has led to the development of printed circuit board (PCB) metasurfaces whose constituent meta-atoms are fabricated by etching designed conductive patterns on dielectric substrates. Due to their compact form factor and their unprecedented ability to precisely control various aspects of the electromagnetic waves, such as the phase~\cite{Chen:PRB2018,Lavigne:TAP2018,Xu:TAP2019,Chen:TAP2020}, amplitude~\cite{Wan:SciRep2016}, polarization~\cite{Niemi:TAP2013,Xu:TAP2016,Kim:PRApplied2020} and frequency content~\cite{Xu:TAP2018_1,Xu:TAP2018_2}, PCB MTSs represent the ideal platform for realizing the next generation of antennas. They have already been successfully integrated into existing technologies such as transmitarray~\cite{Xu:TAP2017,Wu:PRApplied2019} and reflectarray~\cite{Cai:TAP2018,Yang:TAP2018} antennas. 

To further reduce the overall profile of the system without compromising its performance, metasurfaces with integrated sources have been investigated. Periodically modulated impedance MTSs based on the extended principles of holography have been successfully designed to transform  the surface waves from embedded dipoles into directive radiation~\cite{Minatti:TAP2014,Minatti:TAP2016,Ovejero:TAP2017,Faenzi:SciRep2019}. By adding weak perturbations to the average surface impedance, shaped beams can also be produced. Some of the limitations of this approach can be lifted by directly synthesizing the desired aperture fields with the help of local-power-balancing auxiliary surface waves~\cite{Kwon:TAP2020,Kwon:TAP2021}. Alternatively, complete control of the antenna aperture fields can be obtained using HMSs~\cite{Epstein:NatComm2016,Kim:PRApplied2021} or BMSs~\cite{Epstein:TAP2017,Abdo:TAP2019,Ataloglou:TAP2020,Xu:PRA2020} with embedded sources, which offer the designer much more flexibility in terms of the obtainable radiation patterns. However, they are harder to implement and incur more ohmic loss due to their underlying complexity. 

A class of embedded-source-fed passive beam-forming MTSs leveraging non-local electromagnetic interactions have been recently proposed~\cite{Ataloglou:APS2020,Ataloglou:JoM2021,Budhu:MTM2021,Budhu:arxiv2021_3,Xu:APS2021}. In these designs, the feed is placed extremely close to the MTS. The localized source power is distributed across the MTS through tailored auxiliary surface waves (SWs), which enables the effective utilization of the entire physical aperture, regardless of its size. In contrast to holographic modulated impedance MTSs, these devices are capable of truly arbitrary beam-forming, since they do not assume any functional form for the aperture fields. In particular, some of the aforementioned designs~\cite{Budhu:MTM2021,Xu:APS2021} leverage a single electric impedance MTS backed by a ground plane, meaning they are easier to fabricate and can be less lossy than other implementations that rely on multi-layer transmissive HMSs. Their exceptional beam-forming capability can be explained by the fact that the ground plane shorts out the contribution of the effective electric currents to the far-field. Hence, the radiation pattern is solely dictated by the effective magnetic currents, which the electric impedance MTS supplies through its induced conduction currents and images (caused by the ground plane).

Passive and lossless MTSs leveraging auxiliary SWs benefit from a more rigorous design process than conventional ones, due to their reliance on tailored mutual interactions between meta-atoms. To address this, various approaches based on integral equations (IEs) have been developed~\cite{pearson:2020,Brown:TAP2020,Ataloglou:TAP2020,Ataloglou:APS2020,Ataloglou:AWPL2021,Budhu:TAP2021,Brown:TAP2021}. Recently, these methods have been extended to account for the effect of dielectric and/or ground plane truncation~\cite{Budhu:APS2020,Budhu:EUCAP2020,Budhu:EUCAP2021,Budhu:arxiv2021_1,Budhu:arxiv2021_2,Budhu:arxiv2021_3,Xu:APS2021}, which can alter the beam shape through edge reflection and diffraction. Beside their accuracy, another major advantage of IE-based design approaches is that they are valid regardless of the homogeneity of the MTS. Hence, they can be utilized to design devices operating within the refractive (gradient MTS) regime or the diffractive regime. The latter, sometimes referred to as metagratings~\cite{Rabinovich:TAP2018,Rabinovich:TAP2020,Xu:TAP2021_1,Xu:TAP2021_2} or sparse MTSs~\cite{Wong:PRX2018,Popov:PRApplied2020,Popov:AOM2021}, can be desirable due to their low complexity. Furthermore, they are well-suited for synthesizing devices that may have ``forbidden regions" in which meta-atoms cannot be printed~\cite{Salucci:TAP2018}.

In this paper, we present the complete end-to-end design of compact passive and lossless ground-plane-backed impedance MTSs featuring embedded sources. Along with an accurate analysis technique based on a set of coupled volume-surface integral equations (VSIE), we develop an efficient and versatile accompanying optimization-based design method that can realize arbitrary beam-forming through exploitation of tailored auxiliary surface waves. Despite its simplicity, the proposed method can synthesize devices capable of extreme feats such as wide-angle or shared-aperture beam-forming while maintaining near-perfect aperture efficiency. We also derive optimization constraints based on practical considerations. Full-wave simulations with realistic devices confirm that, not only can the synthesized impedance MTSs form arbitrarily shaped beams, but they also exhibit significantly improved bandwidth and power efficiency, owing to the proposed constraints.

%%%%%%%%%%%%%%%%%%%%%%%%%%%%%%%%%%%%%%%%%%%%%%%%%%%%%%
%%%%%%%%%%%%%%%%%%%%%%%%%%%%%%%%%%%%%%%%%%%%%%%%%%%%%%
\section{Theoretical Model}
\label{sec:model}
\subsection{Proposed Architecture}
\label{sec:model:architecture}
\begin{figure}[t]
\centering
\includegraphics[width=0.98\linewidth]{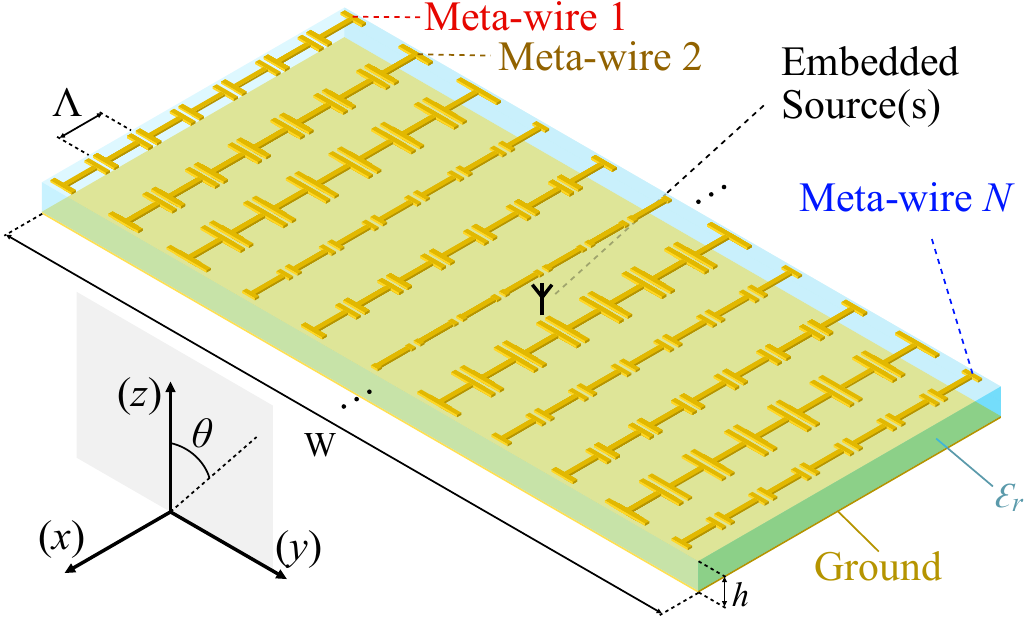}
\caption{Proposed impedance MTS antenna architecture featuring embedded sources.}
\label{fig:MTS_architecture}
\end{figure}

The proposed architecture for the embedded-source-fed impedance metasurface is shown in Fig.~\ref{fig:MTS_architecture}. For simplicity, we restrict our attention to 1D beam-forming devices that are invariant along the $x$-direction. They can be envisioned as collections of ``meta-wires" printed on top of grounded dielectric substrates with thickness $h$, width $W$ (along the $y$-direction), and dielectric constant $\epsilon_r$ . The wires have extremely sub-wavelength loading periodicity ($\Lambda$ in Fig.~\ref{fig:MTS_architecture}), meaning they are homogenizable in their longitudinal ($x$) direction. However, we make no assumption on the homogeneity of the device along the $y$-direction, meaning that the wires can be sparsely and/or non-uniformly spaced. Furthermore, we do not assume an exact form for the embedded source(s) at this stage, other than that it consists of $x$-directed, $x$-invariant electric currents with angular frequency $\omega=2\pi f$. This then implies that the electric field everywhere only contains an $x$-component. Throughout this paper, a time convention of $e^{j\omega t}$ will be used.

\begin{figure}[b]
\centering
\includegraphics[width=0.98\linewidth]{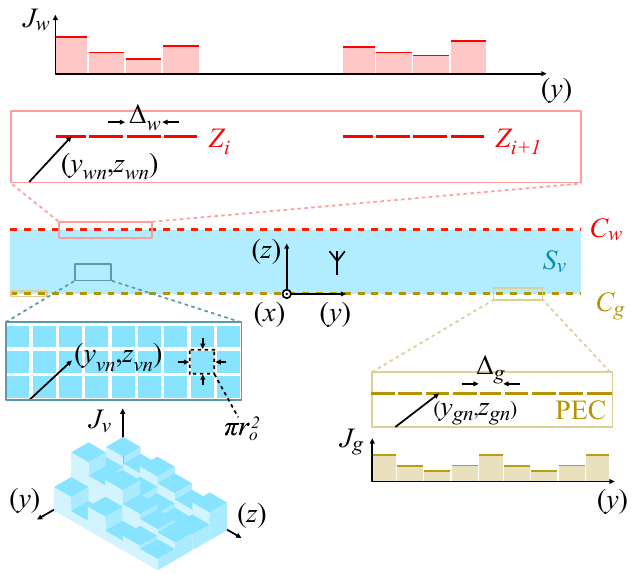}
\caption{Theoretical model for the cross section of the impedance MTS antenna.}
\label{fig:theoretical_model}
\end{figure}

To facilitate the efficient analysis and design of the device under consideration, we model the cross sections of its $N$ meta-wires using hypothetical narrow homogeneous strips residing on the curve $C_w$, as labeled in Fig.~\ref{fig:theoretical_model}. Each strip has a surface electric impedance $Z_i$ ($i\in[1,N]$), which is a function of the printed wire geometry. This is a good approximation as long as the wires have deeply sub-wavelength widths. The impedances of all the wires can be written into a $N\times1$ vector $\bar{Z}_w$. The cross section of the ground plane is modeled by the curve $C_g$, while the cross section of the dielectric is modeled by the rectangular region $S_v$.

\subsection{Volume-Surface Integral Equations Formulation}
\label{sec:model:VSIE}
The 2D model in Fig.~\ref{fig:theoretical_model} allows us to predict and engineer the radiation pattern of the MTS using a modified version of the well known VSIE framework~\cite{Lu:APS1999} originally proposed to model mixed conductive and dielectric structures. It was recently adopted to design metasurface antennas~\cite{Budhu:APS2020,Budhu:EUCAP2020,Budhu:EUCAP2021,Budhu:MTM2021,Budhu:arxiv2021_1,Budhu:arxiv2021_2,Budhu:arxiv2021_3,Xu:APS2021}. For completeness, this approach is reviewed here, along with some of our proposed augmentations.

According to the VSIE formulation, the total field radiated by the MTS antenna, $\vec{E}$,  is ascribed to the following four contributions:

\begin{enumerate}
    \item The source (incident) field $\vec{E}^i$, radiated by the feed as if it was in free space;
    \item The scattered field $\vec{E}^s_g$ radiated by the induced conduction currents $\vec{J}_g$ on the ground plane;
    \item The scattered field $\vec{E}^s_w$ radiated by the induced conduction currents $\vec{J}_w$ on the meta-wires;
    \item The scattered field $\vec{E}^s_v$ radiated by the induced polarization currents $\vec{J}_v$ in the dielectric.
\end{enumerate}

Computation of the total field amounts to solving for the unknown currents induced by $\vec{E}^i$. To that end, a coupled system of integral equations for $\vec{J}_g$, $\vec{J}_w$, $\vec{J}_v$ can be written. It can be solved numerically using the method of moments.

The fields radiated by the unknown surface conduction currents can be written as
\begin{equation}
\label{eqn:scattered_field_conduction}
    E^s_{\iota}\left(\pvec{\rho}\right) = -\frac{k\eta}{4}\int\limits_{C_{\iota}}H^{(2)}_0\left(k\left|\vec{\rho}-\pvec{\rho}'\right|\right)J_\iota(\pvec{\rho}')dy'
\end{equation}
with $\iota\in\left\{g,w\right\}$,  $\pvec{\rho}=\hat{y}y+\hat{z}z$ and $\pvec{\rho}'=\hat{y}y'+\hat{z}z'$. Here, $H^{(2)}_0(\cdot)$ is the zeroth order Hankel function of the second kind, and $k=\omega/c_o$ is the free space wave number.

The fields radiated by the unknown volume polarization current in the dielectric can be written as
\begin{equation}
\label{eqn:scattered_field_polarization}
E^s_{v}\left(\pvec{\rho}\right) = -\frac{k\eta}{4}\iint\limits_{S_v}H^{(2)}_0\left(k\left|\vec{\rho}-\pvec{\rho}'\right|\right)J_v(\pvec{\rho}')dy'dz'.
\end{equation}

The total tangentional field must vanish on the ground plane, meaning
\begin{equation}
\label{eqn:SIE_ground}
E^i+E^s_g+E^s_v+E^s_w = 0 \quad \mathrm{on}~C_g.
\end{equation}

Furthermore, on meta-wire $i$, the total field must be proportional to $J_w$ by $Z_i$, according to Ohm's law. In other words,
\begin{equation}
\label{eqn:SIE_wire}
E^i+E^s_g+E^s_v+E^s_w = Z_iJ_w \quad \textrm{on meta-wire}~i.
\end{equation}

The electric field inside the dielectric must be proportional to the induced polarization current according to~\cite{harrington:1968}
\begin{equation}
\label{eqn:VIE}
E^i+E^s_g+E^s_v+E^s_w = \frac{1}{j\omega(\epsilon_r-1)\epsilon_o}J_v \quad \textrm{in}~S_v.
\end{equation}

Together, (\ref{eqn:SIE_ground})-(\ref{eqn:VIE}) forms a set of VSIEs which can be easily solved using the method of moments.
%to identify the unknown $\vec{J}_g$, $\vec{J}_w$, $\vec{J}_v$ induced by any given $\vec{E}^i$
In this work, we employ the point-matching method, with the discretization scheme illustrated by the zoomed-in pictures in Fig.~\ref{fig:theoretical_model}. The curve representing the ground plane ($C_g$) is divided into $N_g$ segments with width~$\Delta_g$, centered at $\vec{\rho}_{gn}=\hat{y}y_{gn}+\hat{z}z_{gn}$ with $n\in[1,N_g]$. Similarly, each of the impedance strips representing the meta-wires is divided into $N_w'$ contiguous segments with center coordinates $\vec{\rho}_{wn}=\hat{y}y_{wn}+\hat{z}z_{wn}$, yielding a total of $N_w=N_w'\times N$ segments for the entire $N$-wire array. The width of the meta-wire segments are also chosen to be $\Delta_w$. The cross section of the dielectric substrate is discretized into a 2D grid of rectangular (almost square) cells with area $\pi r_o^2$, centered at $\vec{\rho}_{vn}=\hat{y}y_{vn}+\hat{z}z_{vn}$ with $n\in[1,N_v]$. The area is specified this way such that the rectangular cells can be approximated by circles with radius $r_o$ during numerical integration.

As illustrated in Fig.~\ref{fig:theoretical_model}, the described discretization scheme allows one to expand the currents and the fields on $S_g$ and $S_w$ using 1D pulse basis functions. The currents and the fields in $S_v$ can be expanded using 2D pulse basis functions. Then, the VSIEs can be cast into the matrix form

\begin{equation}
\label{eqn:VSIE_matrix}
\begin{split}
        \bar{E}^i=\left(\mathbf{Z}-\mathbf{G}\right)\bar{J}&\triangleq\mathbf{L}\bar{J},\\
\end{split}
\end{equation}
where
\begin{equation}
\label{eqn:VSIE_matrix_definitions}
  \begin{gathered}
        \bar{E}^i = \begin{bmatrix}
                        \bar{E}^i_g\\
                        \bar{E}^i_v\\
                        \bar{E}^i_w\\
                    \end{bmatrix},\quad\bar{J}=\begin{bmatrix}
                        \bar{J}_g\\
                        \bar{J}_v\\
                        \bar{J}_w\\
                    \end{bmatrix},\\
        \mathbf{G}=\begin{bmatrix}
                        \mathbf{G}_{gg}&\mathbf{G}_{gv}&\mathbf{G}_{gw}\\
                        \mathbf{G}_{vg}&\mathbf{G}_{vv}&\mathbf{G}_{vw}\\
                        \mathbf{G}_{wg}&\mathbf{G}_{wv}&\mathbf{G}_{ww}\\
                    \end{bmatrix},\quad\mathbf{Z}=\begin{bmatrix}
                        \mathbf{0}&\mathbf{0}&\mathbf{0}\\
                        \mathbf{0}&\mathbf{P}&\mathbf{0}\\
                        \mathbf{0}&\mathbf{0}&\mathbf{Z}_w\\
                    \end{bmatrix}.
    \end{gathered}
\end{equation}
The vector $\bar{E}^i_{\iota}$ contains the incident (source) field sampled at the discretization points $\vec{\rho}_\iota$, while $\bar{J}_{\iota}$ contains the sampled conduction or polarization currents. The blocks of the matrix $\mathbf{G}$ represent the self and mutual interaction between the various components of the impedance MTS antenna. They can be populated with (\ref{eqn:scattered_field_conduction}) and (\ref{eqn:scattered_field_polarization}). The integration over the ground plane and meta-wire segments can be performed numerically using the midpoint rule~\cite{Balanis:2012}. The integration over dielectric cells can be estimated as integrals over circles with equal area~\cite{Richmond:1965}. The singularities in the integrands associated with the self terms can be treated with the approximations~\cite{Balanis:2012,Richmond:1965}
\begin{equation}
    \mathbf{G}_{\iota\iota}[n][n] = \begin{cases}
-\frac{k\eta\Delta_\iota}{4}\left[1-j\frac{2}{\pi}\log\left(\frac{1.781k\Delta_\iota}{4e}\right)\right] &\iota\in\{g,w\}\\
-\frac{\eta}{2k}\left[kr_oH_1^{(2)}(kr_o)-2j\right] &\iota=v
\end{cases}.
\end{equation}
Recall that $r_o$ is the radius of a circle that has the same area as one rectangular dielectric cell. The matrix $\mathbf{Z}_w$ is diagonal, with entries corresponding to the effective impedances of the meta-wire segments:
\begin{equation}
    \mathbf{Z}_w = \mathrm{diag}\left[\bar{Z}_{w}[1],\cdots,\bar{Z}_{w}[1],\bar{Z}_{w}[2],\cdots,\bar{Z}_{w}[2],\cdots\right].   
\end{equation}
The matrix $\mathbf{P}$ is also a diagonal matrix, whose elements are proportional to the electric susceptibility of the substrate:
\begin{equation}
    \mathbf{P} = \frac{1}{j\omega(\epsilon_r-1)\epsilon_o}\mathbf{1}^{N_v\times N_v}.  
\end{equation}
Here, $\mathbf{1}^{N_v\times N_v}$ is the $N_v\times N_v$ identity matrix.

Equation (\ref{eqn:VSIE_matrix}) can be interpreted as a function for $\bar{J}$ in terms of $\bar{Z}_w$, which can be used to determine the currents induced in the MTS by a known incident field $\bar{E}^i$. The results then enable the evaluation of the total far-field radiated by the MTS through the use of the asymptotic expression of the Hankel function for large arguments~\cite{Balanis:2012}. When observing at some arbitrary distance from the origin (chosen to be 1~m in this study for convenience), and sampled at a set of $N_\theta$ discrete elevation angles, the far-field can be described by the following equation:
\begin{equation}
\label{eqn:radiated_farfield}
\begin{split}
    \bar{E}^{ff}&=\bar{E}^{fi}+[\mathbf{G}_{fg}\quad\mathbf{G}_{fv}\quad\mathbf{G}_{fw}] \bar{J},\\
    \mathbf{G}_{f\iota}[m][n]&=-\frac{k\eta F}{4}\sqrt{\frac{2j}{\pi k}}e^{-jk}e^{jk(y_{\iota n}\sin\theta_m+z_{\iota n}\cos\theta_m)}.\\
\end{split}
\end{equation}
Here, $F=\Delta_\iota$ for $\iota\in\{g,w\}$, and $F=\pi r_o^2$ if $\iota=v$. The $m^{th}$ entry of $\bar{E}^{ff}$ and $\bar{E}^{fi}$ correspond to the total and the source far-field measured at the elevation angle $\theta_m\triangleq2\pi m/N_\theta$.

To complete the analytical model, one can calculate the radiation intensity pattern of the antenna using the total far-zone electric field amplitude, according to
\begin{equation}
\label{eqn:radiation_pattern}
\bar{U}=\frac{1}{2\eta_o}\bar{E}^{ff}\odot\left(\bar{E}^{ff}\right)^*,
\end{equation}
where $\odot$ denotes element-wise product between vectors, and $\{\cdot\}^*$ denotes complex conjugation. This then can be used to calculate various key antenna parameters such as the 2D directivity, defined as
\begin{equation}
    \bar{D} = \frac{N_\theta\bar{U}}{\sum_{n=1}^{N_\theta}\bar{U}(\theta_n)}.
\end{equation}

\subsection{Acceleration With Kron Reduction}
\label{sec:model:acceleration}
In this work, the main purpose for developing the VSIE framework is to enable the optimization of the radiation intensity pattern~$\bar{U}$, with the meta-wire loadings $\bar{Z}_w$ as the tunable variables. Such a procedure necessitates the frequent computation of $\bar{U}$ with updated values of $\bar{Z}_w$. The most direct way to do so is to first find the induced currents in the updated device using (\ref{eqn:VSIE_matrix}), and then compute the new radiation pattern using (\ref{eqn:radiation_pattern}). This  approach is extremely inefficient, since it involves the inversion of a large matrix $\mathbf{L}$, despite the fact that only its smallest block ($\mathbf{Z}_w-\mathbf{G}_{ww}$) is updated. Depending on the size of the antenna and the thickness of the dielectric substrate (which dictate the size of $\mathbf{G}_{vv}$, the largest block of $\mathbf{L}$), the computational burden can be prohibitive. Fortunately, this bottleneck can be resolved with the Kron reduction technique often used in power system analysis~\cite{Grainger:1994}. First, we rewrite (\ref{eqn:VSIE_matrix}) into the following form by regrouping its blocks:
\begin{equation}
\label{eqn:reduced_VSIE_matrix}
        \begin{bmatrix}
        \bar{E}^{i\star}\\
        \bar{E}^i_w\\
    \end{bmatrix} =  \begin{bmatrix}
                        \mathbf{A} & \mathbf{B}\\
                        \mathbf{C} & \mathbf{Z}_w-\mathbf{G}_{ww}\\
                    \end{bmatrix}
                    \begin{bmatrix}
                        \bar{J}^\star\\
                        \bar{J}_w\\
                    \end{bmatrix}.
\end{equation}
Rearranging (\ref{eqn:reduced_VSIE_matrix}) further, we obtain the Kron reduced system
\begin{equation}
\label{eqn:Schur_Jw}
    \bar{J}_w = \mathbf{S}^{-1}\left(\bar{E}^{i}_w-\mathbf{C}\mathbf{A}^{-1}\bar{E}^{i\star}\right),
\end{equation}
where $\mathbf{S}\in\mathbb{C}^{N_w\times N_w}$ is the Schur complement of $\mathbf{A}$, given by
\begin{equation}
\label{eqn:S_definition}
    \mathbf{S}\triangleq \mathbf{Z}_w-\mathbf{G}_{ww}-\mathbf{CA}^{-1}\mathbf{B}.
\end{equation}
Additionally,
\begin{equation}
\begin{split}
\label{eqn:Schur_Jstar}
    \bar{J}^{\star}&\triangleq\begin{bmatrix}
                    \bar{J}_g\\ \bar{J}_v\\
    \end{bmatrix}\\
    &=\left(\mathbf{A}^{-1}+\mathbf{A}^{-1}\mathbf{BS}^{-1}\mathbf{CA}^{-1}\right)\bar{E}^{i\star}-\mathbf{A}^{-1}\mathbf{BS}^{-1}\bar{E}^i_w.
\end{split}
\end{equation}
The far-field can be rewritten as
\begin{equation}
\label{eqn:Schur_Eff}
\begin{split}
    \bar{E}^{ff}=&\left(\mathbf{G}_{fw}-[\mathbf{G}_{fg}\quad\mathbf{G}_{fv}]\mathbf{A}^{-1}\mathbf{B}\right)\bar{J}_w\\
    &+I_o\bar{G}_{fi}+[\bar{G}_{fs}\,\,\bar{G}_{fv}]\mathbf{A}^{-1}\bar{E}^\star.
\end{split}
\end{equation}

Evidently, each evaluation of the cost function using the reduced system requires only the inversion of a single $N_w\times N_w$ matrix, $\mathbf{S}$. The only other matrix inverse ($\mathbf{A}^{-1}$) is static throughout the entire optimization process, because it does not contain~$\mathbf{Z}_w$. Therefore, it can be stored and reused. Furthermore, (\ref{eqn:Schur_Jstar}) accelerates the far-field computation by reducing the size of the matrices being multiplied. In fact, it eliminates the need to explicitly evaluate $\bar{J}^\star$, which contains the currents on the ground plane and in the dielectric.

With the method proposed in this section, the optimization process becomes extremely efficient, regardless of the exact algorithm used. This opens up the opportunity to explore more computationally expensive methods such as global optimization, which may yield better performing designs. Furthermore, one can discretize the dielectric substrate and the ground plane with very fine resolution in order to improve the accuracy of the results, without incurring too much additional computational cost during optimization.

%%%%%%%%%%%%%%%%%%%%%%%%%%%%%%%%%%%%%%%%%%%%%%%%%%%%%%
%%%%%%%%%%%%%%%%%%%%%%%%%%%%%%%%%%%%%%%%%%%%%%%%%%%%%%
\section{Far-field Optimization}
\label{sec:optimization}
The closed-form expressions presented in Sec.~\ref{sec:model:VSIE} as well as the acceleration method proposed in Sec.~\ref{sec:model:acceleration} enable the rapid optimization of antenna characteristics. In this work, as a proof of concept, we demonstrate the ability for the compact MTS system to realize arbitrary beam-forming by shaping $\bar{U}(\bar{Z}_w)$ to match stipulated radiation patterns.
\subsection{Optimization method}
Different optimization methods can be used to determine the effective impedances of the meta-wires $\bar{Z}_w$ that produce a desired far-field radiation, based on \eqref{eqn:radiation_pattern}, \eqref{eqn:Schur_Jw}, \eqref{eqn:S_definition} and \eqref{eqn:Schur_Eff}. For the examples presented in this work, we mainly rely on gradient-descent optimization, as implemented by the built-in function \textit{fmincon} in MATLAB. This function is capable of locally minimizing a cost function while adhering to multiple linear and/or nonlinear constraints on the solution.

In order to provide a starting point $\bar{Z}_w^{(0)}$ in the gradient-descent optimization, a preliminary step needs to be performed. This preparatory step helped the optimizer converge to solutions with better pattern matching in the vast majority of the studied design examples. Specifically, following previously proposed methods~\cite{Budhu:TAP2021,Budhu:arxiv2021_1,Budhu:arxiv2021_2}, we try to determine the currents $\bar{J}_w$ that produce the desired far-field radiation, based on \eqref{eqn:radiation_pattern} and \eqref{eqn:Schur_Eff}. We do this with an unconstrained gradient-descent optimization using the function \textit{fminunc} in MATLAB. Then, a set of complex loadings that would give rise to the optimized $\bar{J}_w$ is calculated based on Ohm's law in \eqref{eqn:SIE_wire}. The real part of the impedances is discarded and the imaginary part is used as the starting point $\bar{Z}_w^{(0)}$ of the main gradient-descent optimization~\cite{Budhu:TAP2021,Budhu:arxiv2021_1,Budhu:arxiv2021_2}.

For some designs that have more demanding specifications, this method may not be able to supply a good starting $\bar{Z}_w^{(0)}$. Such scenarios demand more specialized treatments, which will be discussed as they arise.

\subsection{Cost Function}
\label{sec:optimization:cost_function}
The aim of the optimization is to match the radiation pattern $\bar{U}(\theta_m)$ produced from the metasurface with a desired radiation pattern $\bar{U}_\mathrm{des}(\theta_m)$. Since it is not clear what the total power contained in $\bar{U}_\mathrm{des}(\theta_m)$ should be, we choose to define the cost function in terms of normalized radiation patterns, as
\begin{align}
\label{eqn:pattern_matching_cost_function}
    F = \sum_{m=1}^{N_\theta} \left( \frac{\bar{U}(\theta_m)}{\max_{\theta_m}\{\bar{U}(\theta_m)\}}- \frac{\bar{U}_\mathrm{des}(\theta_m)}{\max_{\theta_m}\{\bar{U}_\mathrm{des}(\theta_m)\}} \right)^2.
\end{align}
Due to the normalization, (\ref{eqn:pattern_matching_cost_function}) can also be interpreted as a measure of the difference between the desired directivity pattern $\bar{D}_{des}$ and the realized pattern $\bar{D}$.

It should be noted that choosing to form the cost function this way, as we have done in this study, clearly puts the focus on the main beam compared to the side lobe values. However, even the shape of the side lobes will eventually match the expected one, if the algorithm converges to a low value of~$F$.

The choice to match the normalized radiation pattern also means that the converged solution may not have the highest possible radiation efficiency. Indeed, there can be an infinite set of solutions satisfying the same constraints, but having different capacities for near-field reactive energy storage. There are multiple methods to deal with poorly radiating solutions. One way is to design a simple matching network for the input. Alternatively, in Sec.~{\ref{sec:optimization:constraints}}, we introduce a constraint which can help guide the optimizer towards solutions with inherently high radiation efficiency.

\subsection{Optimization constraints}
\label{sec:optimization:constraints}
In order for the optimizer to converge to solutions with desired characteristics, we introduce a set of linear and nonlinear constraints on the wire impedances. First, we require that the metasurface is passive and lossless. Therefore, the impedances should be purely imaginary, i.e. $\mathrm{Re} \{Z_n \} =0, \forall n$. This constraint implies that the optimized device does not rely on engineered ohmic loss or power gain, both of which are difficult to control in practice. We also demand that the impedances be practically realizable with printed, periodically loaded copper traces which adhere to standard fabrication tolerances. By varying the width of a printed loading capacitor, an impedance range of $[-j90,-j25]$ was acquired, as described in Appendix~\ref{append:UnitCell}. It is noted that interdigitated capacitors or other types of scatterers can increase this range. However, no significant improvements were observed in terms of the pattern matching for the designs considered in this paper. Therefore, the aforementioned impedance range was deemed sufficient.

A few more optional constraints can be introduced to converge to solutions with higher bandwidth and lower sensitivity. Having the currents $\bar{J}_w$ at each iteration allows us to calculate the fields anywhere in the near-field region of the metasurface (including in the dielectric), as well as the radiated power. In general, it is advantageous to have as much radiated power as possible, while maintaining the field amplitudes and the stored energy close to the antenna at a relatively low level. With this in mind, we place a constraint on the quality factor (Q-factor) of the structure, calculated as:
\begin{equation}
\label{eqn:Qconstraint}
    Q\triangleq\omega_0 \frac{W}{P_\mathrm{rad}} < Q_\mathrm{max},
\end{equation}
where 
\begin{equation}
\label{eqn:stored_energy}
    W=\frac{1}{2} \iint \epsilon_0 \epsilon_r (\vec{\rho}) |\mathbf{E}(\vec{\rho})|^2 dydz
\end{equation} 
serves as an estimate of the stored energy and $P_\mathrm{rad}$ is the total radiated power as calculated from the radiation intensity $\bar{U}$. It is noted that the integration in (\ref{eqn:stored_energy}) extends beyond the dielectric region, into the air region close to the metasurface. This allows for the capture of the confined surface waves which also contribute to energy storage.

Lastly, we can put a constraint on the sensitivity of the far-field with respect to the wire impedances, by computing the derivative $d\bar{E}^{ff}/dZ_n$. This constraint helps the optimizer to converge to solutions which are more robust against fabrication errors that may modify the effective wire impedances. Moreover, such a trait also relates to the frequency bandwidth of the device, since the effective impedances for realistic meta-wires are frequency-dispersive. As seen from \eqref{eqn:Schur_Jw} and \eqref{eqn:Schur_Eff}, to calculate the derivative of the far-field with respect to each loading $Z_n$, we basically have to calculate the term $d\mathbf{S}^{-1}/dZ_n$. This can be done analytically without any extra matrix inversion, by using the identity for the derivative of an inverse matrix:
\begin{equation}
\frac{d\mathbf{S}^{-1}}{dZ_n}=-\mathbf{S}^{-1}\left(\frac{d\mathbf{S}}{dZ_n}\right)\mathbf{S}^{-1}.
\end{equation}
The inverse $\mathbf{S}^{-1}$ is already calculated in each iteration, while the remaining derivative $(d\mathbf{S}/dZ_n)$ reduces, from \eqref{eqn:S_definition}, to a diagonal matrix with $N_w'$ non-zero entries (of $-j$) corresponding to the $n^{th}$ wire. This enables rapid calculation of the derivative of each far-field angle with respect to each wire. 

Having derived $d\bar{E}^{ff}(\theta_m)/dZ_n$, the constraint on sensitivity can be stated as
\begin{align}
\label{eqn:SensitivityConstraint}
    T\triangleq\max_n \left(\sum_{\theta_m} \left|\frac{dE^{ff}(\theta_m)}{dZ_n}\right|\right) < T_\mathrm{max}.\\
\end{align}

As a strategy, the constraints on the passivity ($\mathrm{Re} \{Z_n \} =0$) and the impedance range are always present during optimization. The other two nonlinear constraints, if desired, can be introduced in an additional step, after an initial round of optimization without them. The converged result of the first step is fed to the second, more stringent step, as the new starting point.

%%%%%%%%%%%%%%%%%%%%%%%%%%%%%%%%%%%%%%%%%%%%%%%%%%%%%%
%%%%%%%%%%%%%%%%%%%%%%%%%%%%%%%%%%%%%%%%%%%%%%%%%%%%%%
\section{Numerical Results}
\label{sec:numerical_results}
We verify the proposed embedded-source-fed impedance MTS and the developed design scheme with several numerical examples. The optimized devices are simulated in Ansys HFSS. Since they are $x$-invariant, we can predict their performance by taking a subwavelength-thin slice and placing it inside a parallel plate waveguide with perfectly matched terminations on all open sides. In this section, for simplicity, the meta-wires are modeled using thin strips with uniform surface impedances. We consider realistic devices with printed metallic wires in Sec.~\ref{sec:results:realistic}.

\subsection{Wide-angle Beam-forming with Near-perfect Aperture Efficiency}
\label{sec:numerical_results:uniform_aperture}
In this subsection, we demonstrate the ability of the proposed MTS topology to perform wide-angle beam-forming with near unity aperture efficiency. The size of the device is $7\lambda$ at the design frequency of 10~GHz. The impedance MTS on top of the grounded substrate ($h=2.54$~mm, $\epsilon_r=3$) only has 28 meta-wires, leading to a relatively large inter-element spacing of $\lambda/4$. Each meta-wire is modeled using a narrow impedance strip with width $0.7$~mm. The embedded source is chosen to be a single uniform line current located at \mbox{$(y_o,z_o)=(0,h/2)$}. 

The proposed device can be seen as a simplified version of the previously presented cavity-excited antennas~\cite{Epstein:NatComm2016,Epstein:TAP2017}, featuring a dramatically reduced profile as well as a less complicated and sparser metasurface layer. Furthermore, the tasks of source power redistribution and aperture efficiency optimization, previously performed by a resonant cavity mode, is now accomplished by an auxiliary surface wave on the MTS. Despite the different working principle, the proposed MTS is still able to realize near-perfect illumination of arbitrarily large apertures with a single localized source~\cite{Budhu:MTM2021,Xu:APS2021}.

To begin the design, we first note that the incident field is simply a cylindrical wave centered at the source location $\vec{\rho}_o=\hat{y}y_o+\hat{z}z_o$. Hence the incident near-field vectors $\bar{E}^i_\iota$ can be easily populated following
\begin{equation}
    \label{eqn:line_source_incident_field}
    \bar{E}^i_{\iota}[n]=-\frac{k\eta I_o}{4}H_0^{(2)}\left(k\left|\vec{\rho}_{\iota n}-\vec{\rho}_o\right|\right),\quad \iota\in\{g,v,w\},
\end{equation}
where $I_o$ is an arbitrarily chosen source current amplitude. The far-field radiated by the source is given by
\begin{equation}
    \bar{E}^{fi}[n]=-I_o\frac{k\eta}{4}\sqrt{\frac{2j}{\pi k}}e^{-jk}e^{jk(y_o\sin\theta_n+z_o\cos\theta_n)}\quad \forall n.
\end{equation}

It is desired to form a directive beam towards some arbitrary direction $\theta_o$ with the highest possible aperture efficiency. Hence, the target far-zone electric field is that associated with a uniform aperture with linear phase gradient $k\sin\theta_o$. It can be interpreted as the fields radiated by a sheet of phased surface current with uniform amplitude, given by
\begin{equation}
    \label{eqn:uniform_aperture_radiation}
    \begin{split}
        \bar{E}^{ff}_{des} &= \mathbf{G}_{fg}\bar{J}_{des},\\
        \bar{J}_{des}[n] &= J_oe^{-jky_{gn}\sin\theta_o},\\
    \end{split}
\end{equation}
where $J_o$ is some arbitrary complex constant which will be dropped upon normalization in (\ref{eqn:pattern_matching_cost_function}). It is chosen to be 1 [A/m] in this work.

\begin{figure}[t]
\centering
\includegraphics[width=0.98\linewidth]{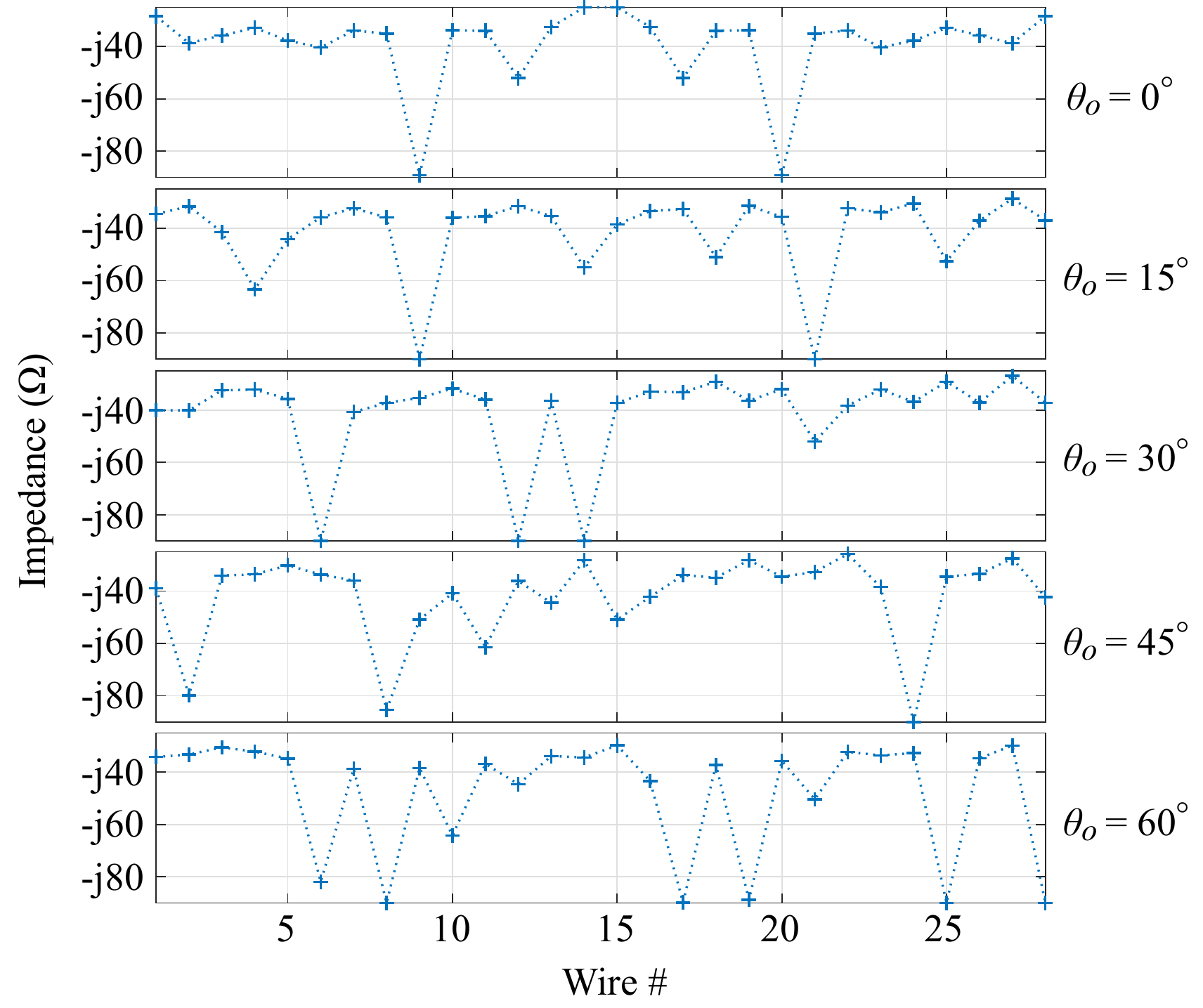}
\caption{Optimized meta-wire impedances for wide-angle beam-forming with perfect aperture efficiency.}
\label{fig:uniform_aperture_impedance}
\end{figure}

We optimize the meta-wire impedances to produce beams with output angles ranging from $\theta_o=-60^\circ$ to $\theta_o=0^\circ$ with $15^\circ$ increments. The converged impedance values are plotted in Fig.~\ref{fig:uniform_aperture_impedance}. Importantly, due to the acceleration discussed in Sec.~\ref{sec:model:acceleration} and the sparsity of the MTS, each design only took about 20 seconds to converge to the optimum.

\begin{figure}[b]
\centering
\includegraphics[width=0.98\linewidth]{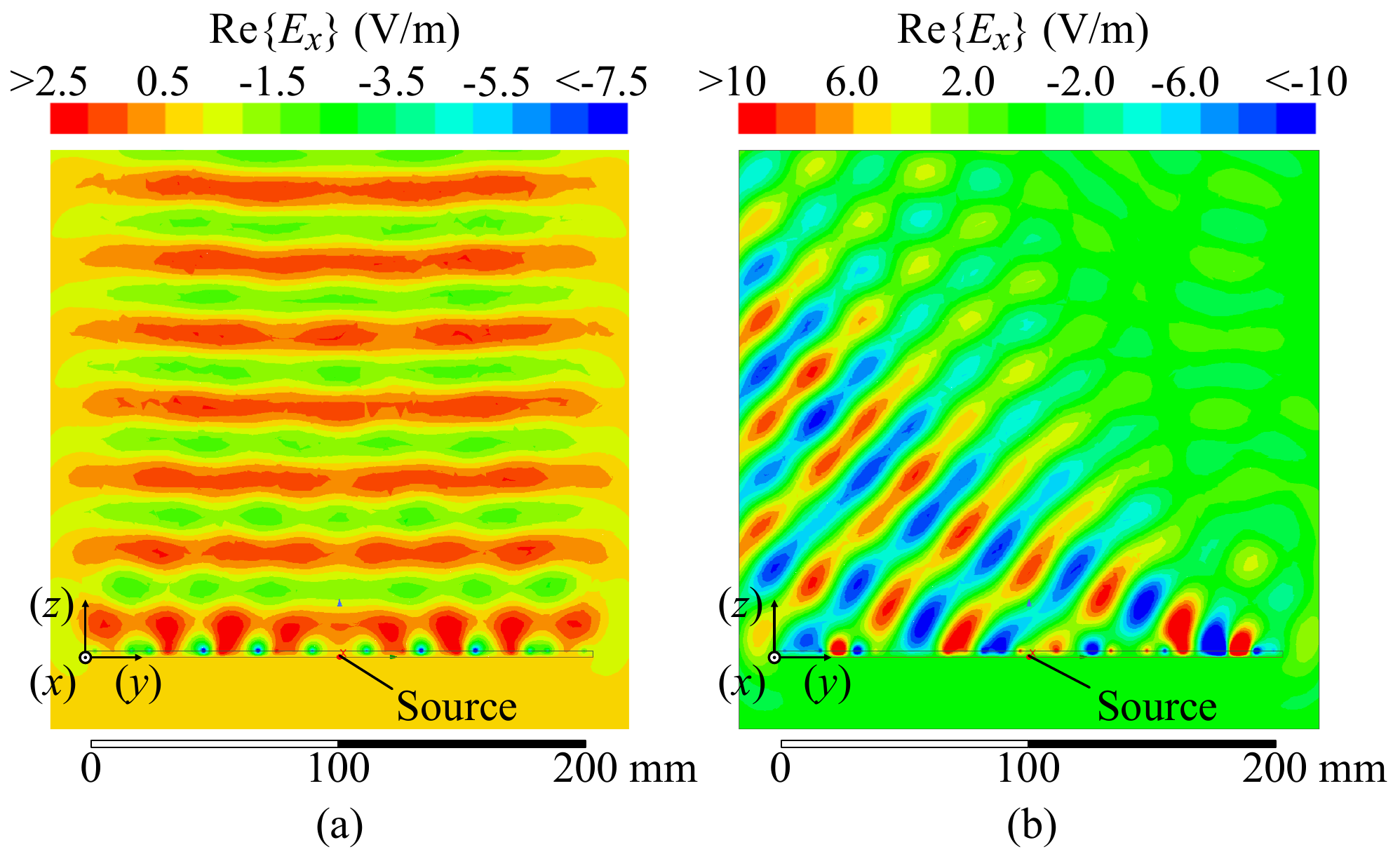}
\caption{Simulated near-field electric field distributions corresponding to (a) $\theta_o=0^\circ$ and (b) $\theta_o=-45^\circ$.}
\label{fig:uniform_aperture_NF}
\end{figure}

The performance of the optimized designs is qualitatively demonstrated by the simulated near-field electric field distributions plotted in Fig.~\ref{fig:uniform_aperture_NF}(a) and (b), which showcase the $\theta_o=0^\circ$ and $\theta_o=-45^\circ$ cases respectively. Here, it can be seen that full utilization of the entire aperture is achieved despite its large electrical size and the poor inherent illumination provided by the embedded source.

\begin{figure}[t]
\centering
\includegraphics[width=0.98\linewidth]{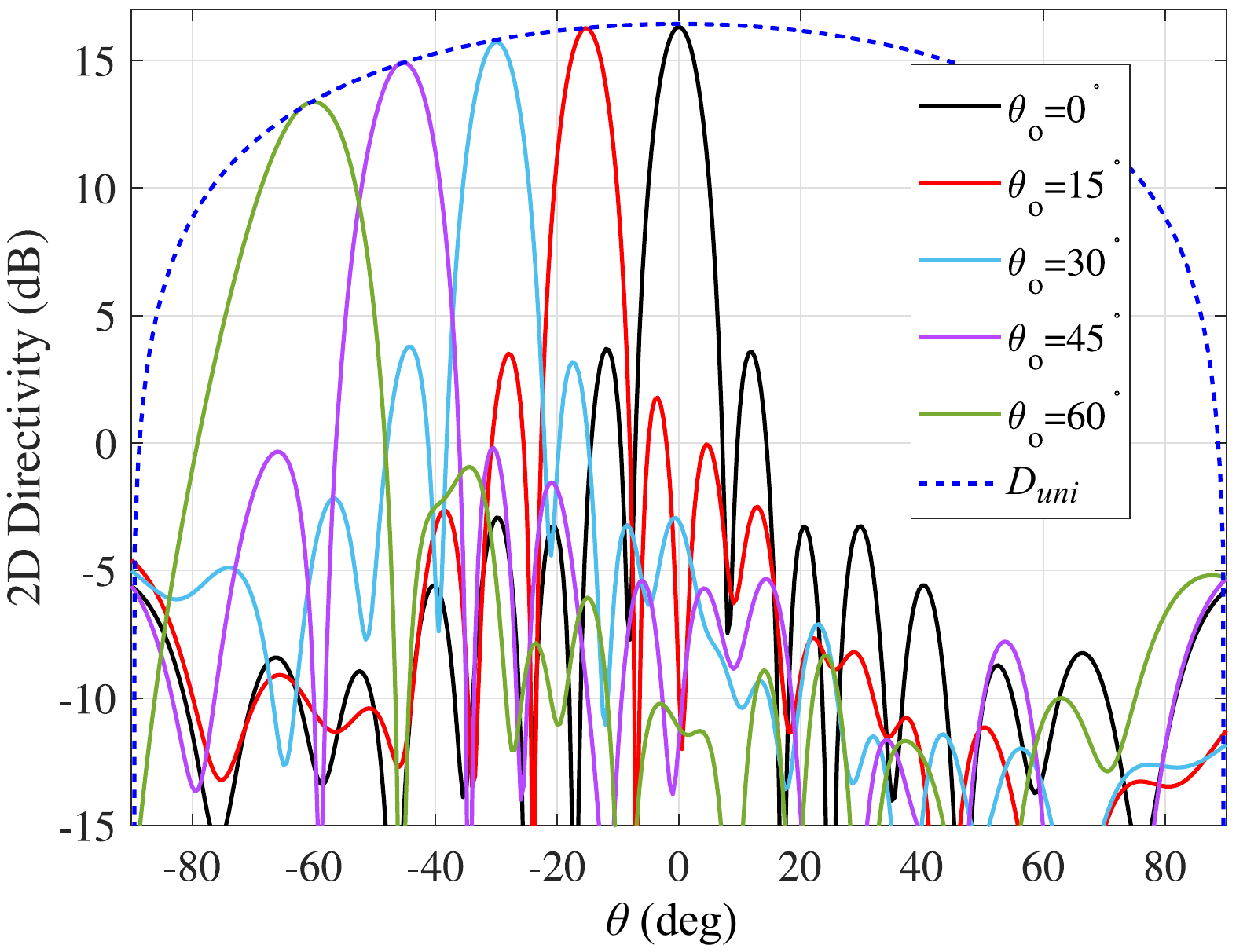}
\caption{Simulated 2D directivity for various designs demonstrating wide-angle beam-forming capability.}
\label{fig:uniform_aperture_dir}
\end{figure}

A more quantitative study on the MTS designs is conducted through an examination of their simulated 2D directivity $D(\theta)$, which are plotted in Fig.~\ref{fig:uniform_aperture_dir}. Here it is shown that the internally excited MTS can easily form highly directive beams up to $-60^\circ$ while maintaining side lobe levels (SLLs) below $-14$~dB. Since the source is placed at the center of the device, it is evident that the exceptional beam-forming ability can be extended to positive values of $\theta$.

The dotted blue curve in Fig.~\ref{fig:uniform_aperture_dir}, labelled as ``$D_{uni}$", represents the 2D directivity of a uniformly illuminated aperture ($W=7\lambda$) radiating in the direction of $\theta$, which is equal to
\begin{equation}
\label{eqn:uniform_directivity}
    D_{uni}(\theta)= \frac{2\pi W}{\lambda}\cos\theta.
\end{equation}
Since the directivity of all simulated beams follow this envelope exactly, one can conclude that each of the designs presented in this section exhibits near-unity aperture efficiency ($\xi_{apt}$), which is defined as
\begin{equation}
\label{eqn:aperture_efficiency}
    \xi_{apt}\triangleq\frac{D(\theta_o)}{D_{uni}(\theta_o)}.
\end{equation}
Calculations using (\ref{eqn:aperture_efficiency}) reveal that each of the presented designs indeed has an aperture efficiency of at least 99\%.

\begin{figure}[b]
\centering
\includegraphics[width=0.98\linewidth]{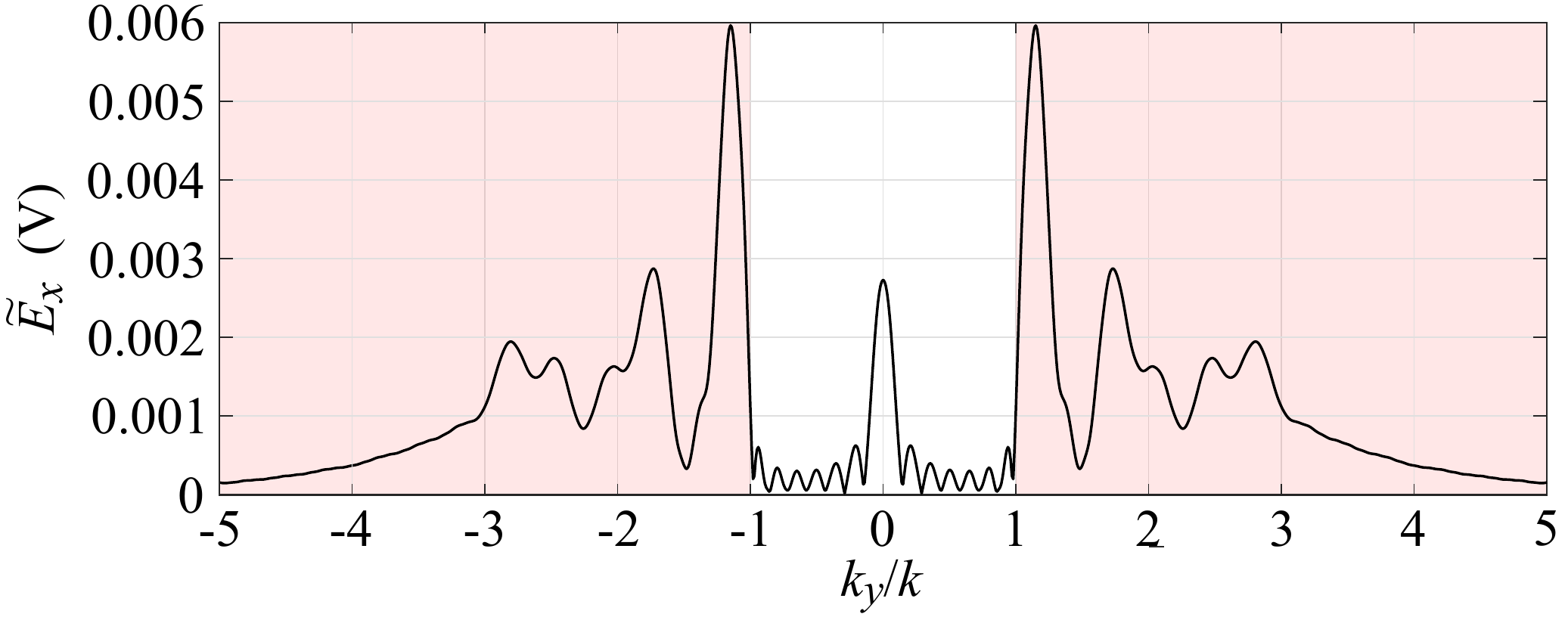}
\caption{Electric field spectrum for the impedance MTS antenna with $\theta_o=0^\circ$, evaluated at a plane $0.1\lambda$ above the meta-wires. The invisible region is shaded in red.}
\label{fig:uniform_aperture_spectrum}
\end{figure}

To verify the contribution of surface waves to the performance of the impedance MTS antennas, we plot the Fourier spectrum of the electric field ($\tilde{E}_x(k_y)$) for the \mbox{$\theta_o=0^\circ$} design in Fig.~\ref{fig:uniform_aperture_spectrum}. The Fourier transform was performed in a plane 0.1$\lambda$ above the meta-wires. The red shaded region in Fig.~\ref{fig:uniform_aperture_spectrum} corresponds to the invisible region (\mbox{$|k_y|>k$}), in which the surface waves reside. Although the spectral content in this region does not directly shape the far-field radiation pattern, they exert an indirect influence by modifying the aperture field distribution. As seen by the high amplitude spectra inside the invisible region in Fig.~\ref{fig:uniform_aperture_spectrum}, the optimized design relies on surface waves to achieve its near perfect aperture illumination.

\subsection{Multi-input Multi-output Metasurfaces}
\label{sec:numerical_results:MIMO}
In this section, we demonstrate a multi-input multi-output (MIMO) beam-forming MTS which is capable of generating two independent beams when one of its two designated embedded sources are excited. Unlike single-input multi-beam systems, the MIMO MTSs are capable of distinguishing the signals picked up by different beams~\cite{Ovejero:TAP2017}. This feature makes them ideal for realizing high-capacity communication links. 

For illustrative purposes, we focus our attention on a dual-input dual-output system. Since the two output beams must share the same physical aperture, it is expected that they each suffer from degraded directivity. However, as we will show, it is possible to obtain significantly higher aperture efficiency for each beam with the proposed optimization-based method, as compared to that achievable with physical aperture partitioning.

The design of the MIMO MTS requires some simple reformulation of the optimization problem. All derived formulae as well as the Kron reduction technique discussed in Sec.~\ref{sec:model} are still valid. However, we now have a set of two independent input fields $\bar{E}^i_{\{1,2\}}$, which give rise to two different output radiation patterns $\bar{U}_{\{1,2\}}$. If we denote the desired output pattern corresponding to input $\{1,2\}$ as $\bar{U}_{des,\{1,2\}}$, then a new cost function can be formulated as

\begin{align}
\label{eqn:pattern_matching_cost_function_MIMO}
    F_{M} = \sum_{i=1}^2\alpha_i\sum_{m=1}^{N_\theta} \left( \frac{\bar{U}_i[m]}{\max_{m}\{\bar{U_i}[m]\}}- \frac{\bar{U}_\mathrm{des,i}[m]}{\max_{m}\{\bar{U}_\mathrm{des,i}[m]\}} \right)^2.
\end{align}
Here, $\alpha_i$ is a weight that allows the designer to place emphasis on pattern matching for one of the two input-output (I/O) pairs. In this study, we use equal weight for both I/O pairs.

The main difficulty for the present optimization problem is the identification of a good starting point. A reasonable strategy is to first treat the MIMO MTS as a single-input single-output device whose input is $\bar{E}^i_1+\bar{E}^i_2$, and whose desired output is $\bar{U}_{des,1}+\bar{U}_{des,2}$. This enables the use of the strategy discussed in Sec.~\ref{sec:optimization} to obtain a rational starting point. However, we found that this method, while serviceable, does not guarantee the best results. On the other hand, a two-step optimization process, consisting of a global optimization which generates a starting point for a subsequent local search, produced much better performing designs. In this study, the global optimization is performed using the built-in particle swarm optimization (PSO) routine in MATLAB, while the local search is done with gradient descent ($fmincon$).

To demonstrate the effectiveness of the proposed approach, we design a $7\lambda$-wide impedance MTS ($h=2.54~\mathrm{mm}, \epsilon_r=3$) consisting of 42 meta-wires (each $\lambda/40$ wide), operating at 10~GHz. The sparse $\lambda/6$ spacing between meta-wires means that the device can be easily implemented using realistic PCB MTSs. The two inputs are assumed to be two line sources located at $(y_o,z_o)=(\pm3\lambda,h/2)$. The corresponding desired outputs are two beams directed at $\pm20^\circ$ with 100\% aperture efficiency. Their far-zone electric fields can be obtained via~(\ref{eqn:uniform_aperture_radiation}).

\begin{figure}[t]
\centering
\includegraphics[width=0.98\linewidth]{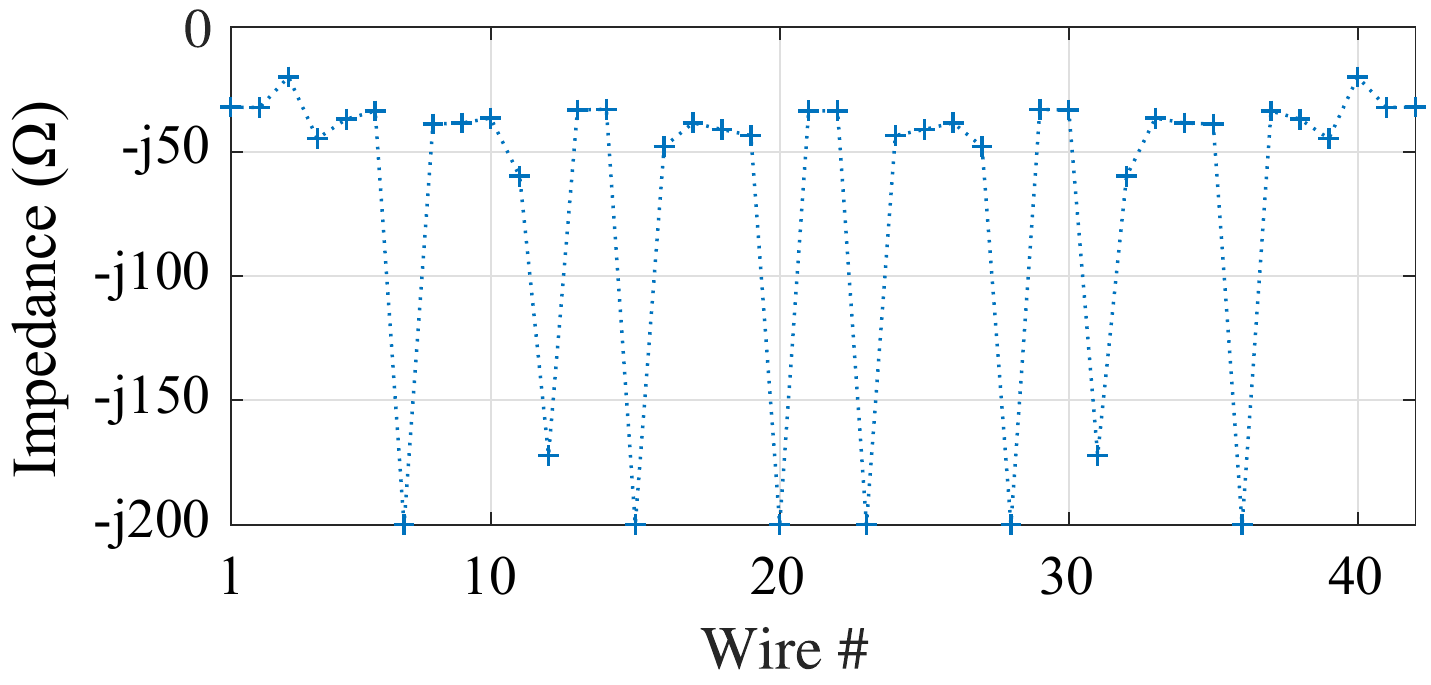}
\caption{Optimized meta-wire reactance for the MIMO MTS.}
\label{fig:MIMO_reactance}
\end{figure}

Since the device is symmetric, only 21 optimization variables need to be considered. To accommodate the more demanding design specifications, we extend the impedance limit of the meta-wires to $[-j200, -j25]$. This range is still easily obtainable using physical meta-wires, as one simply needs to increase the gap between the printed capacitor plates in order to realize more negative values of reactance. Fig.~\ref{fig:MIMO_reactance} depicts the converged $\bar{Z}_w$ after one round of PSO with a swarm size of 50 and an inertia range of $[0.1, 0.5]$, followed by a gradient descent search.

\begin{figure}[b]
\centering
\includegraphics[width=0.98\linewidth]{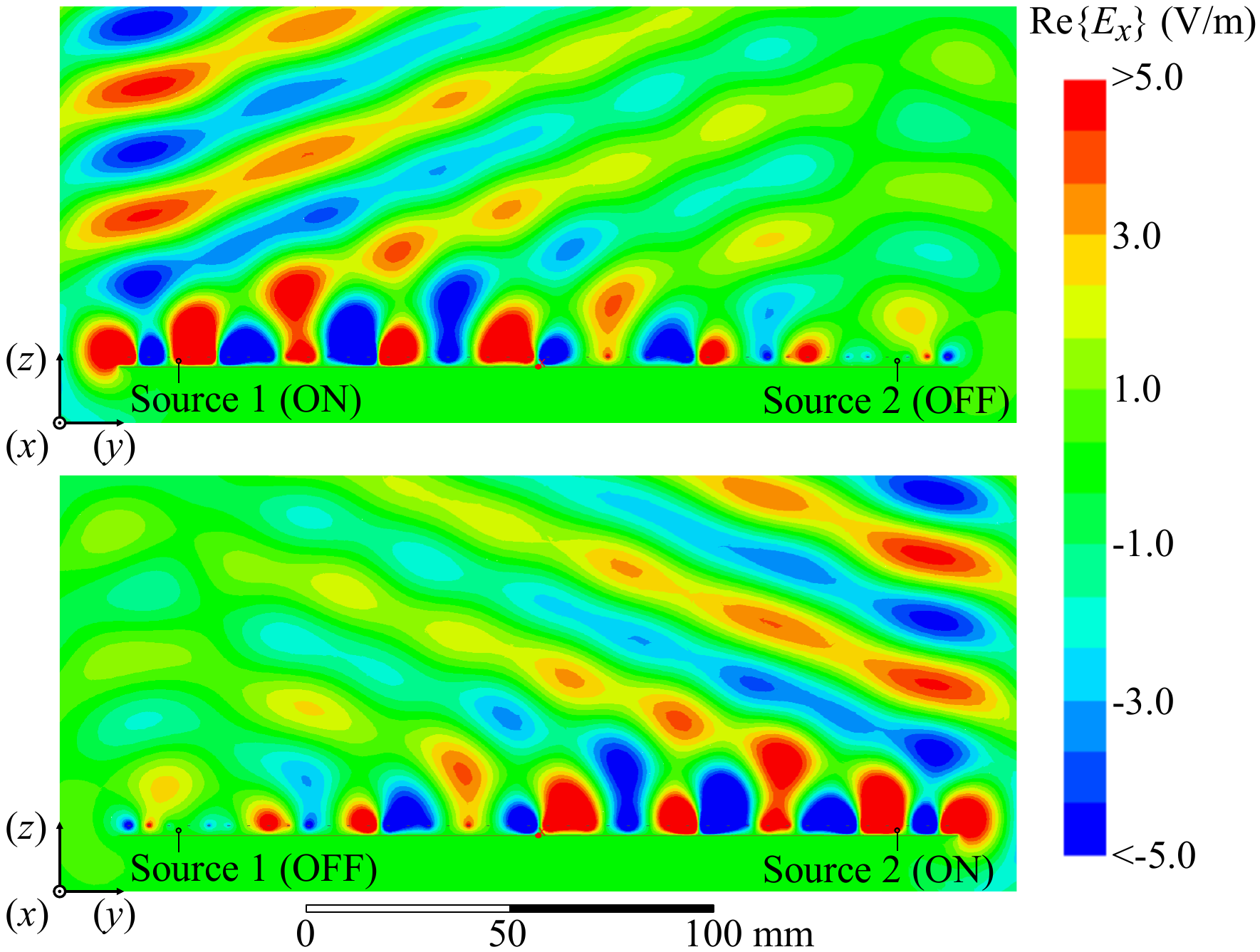}
\caption{Simulated near-field electric field of the MIMO MTS when one of its two inputs is excited.}
\label{fig:MIMO_NF_final}
\end{figure}

The synthesized device is simulated in HFSS, giving the near-field distributions in Fig.~\ref{fig:MIMO_NF_final}. Here, the two subfigures depict the fields radiated by the MTS when one of its two designated inputs is excited. It appears that the two outputs, despite having to share the same physical aperture, are each able to achieve almost full effective aperture utilization. 

\begin{figure}[t]
\centering
\includegraphics[width=0.98\linewidth]{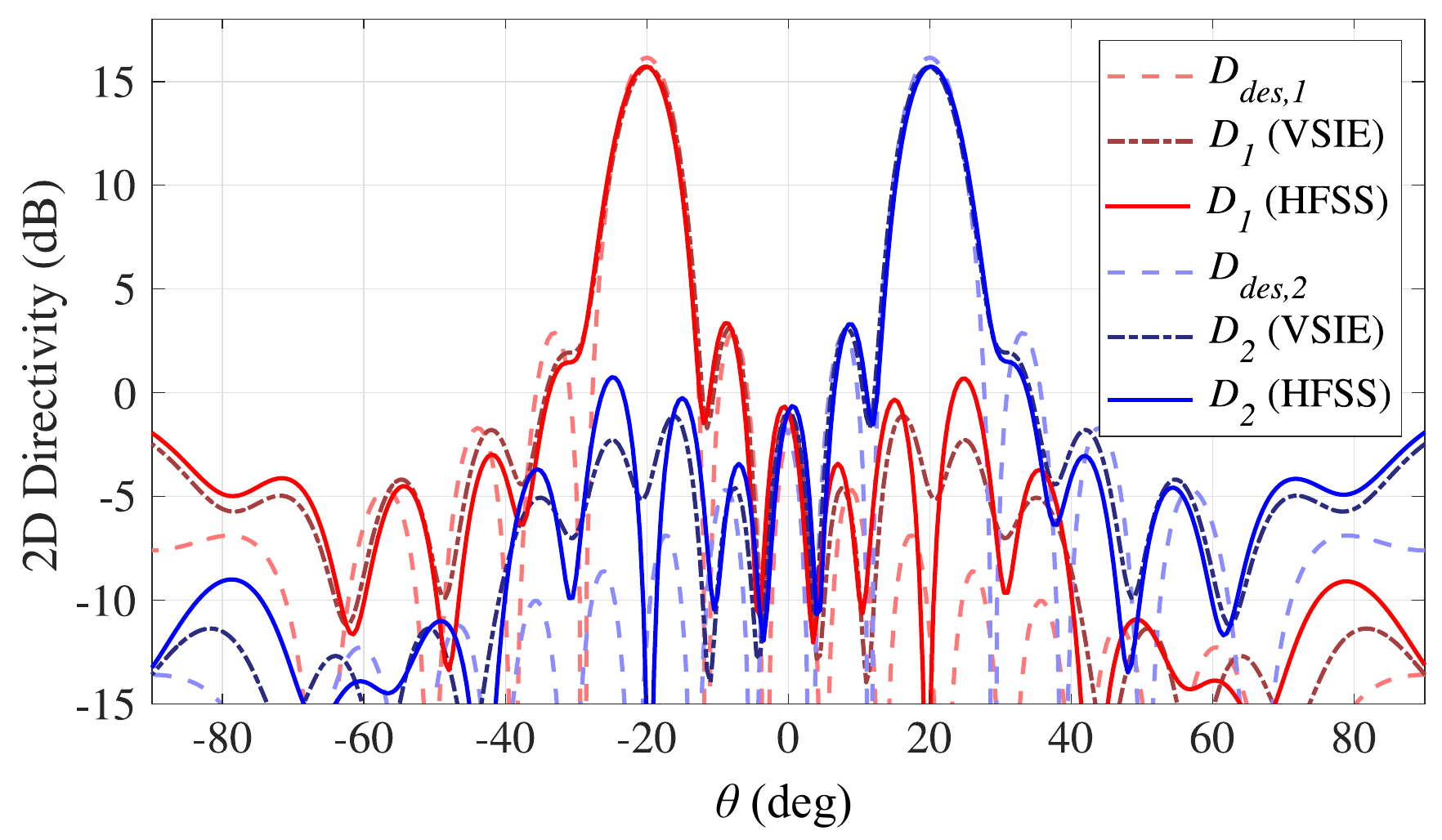}
\caption{Output directivity of the MIMO MTS when one of its two inputs is excited.}
\label{fig:MIMO_dir}
\end{figure}

The above assertion is quantitatively verified via an examination of the simulated 2D directivity patterns, plotted in Fig.~\ref{fig:MIMO_dir}. Here, the dashed red and blue curves correspond to the desired beams produced by uniform apertures. The solid red (blue) curve shows the realized directivity of the MTS when the line source positioned at $-3\lambda$ ($+3\lambda$) is excited. Both output beams exhibit only a 0.47dB drop in directivity when compared to the desired beams, indicating aperture efficiencies of approximately 90\%. This is significantly higher than the efficiency obtainable though heuristic techniques such as physical aperture partitioning.

\begin{figure}[b]
\centering
\includegraphics[width=0.98\linewidth]{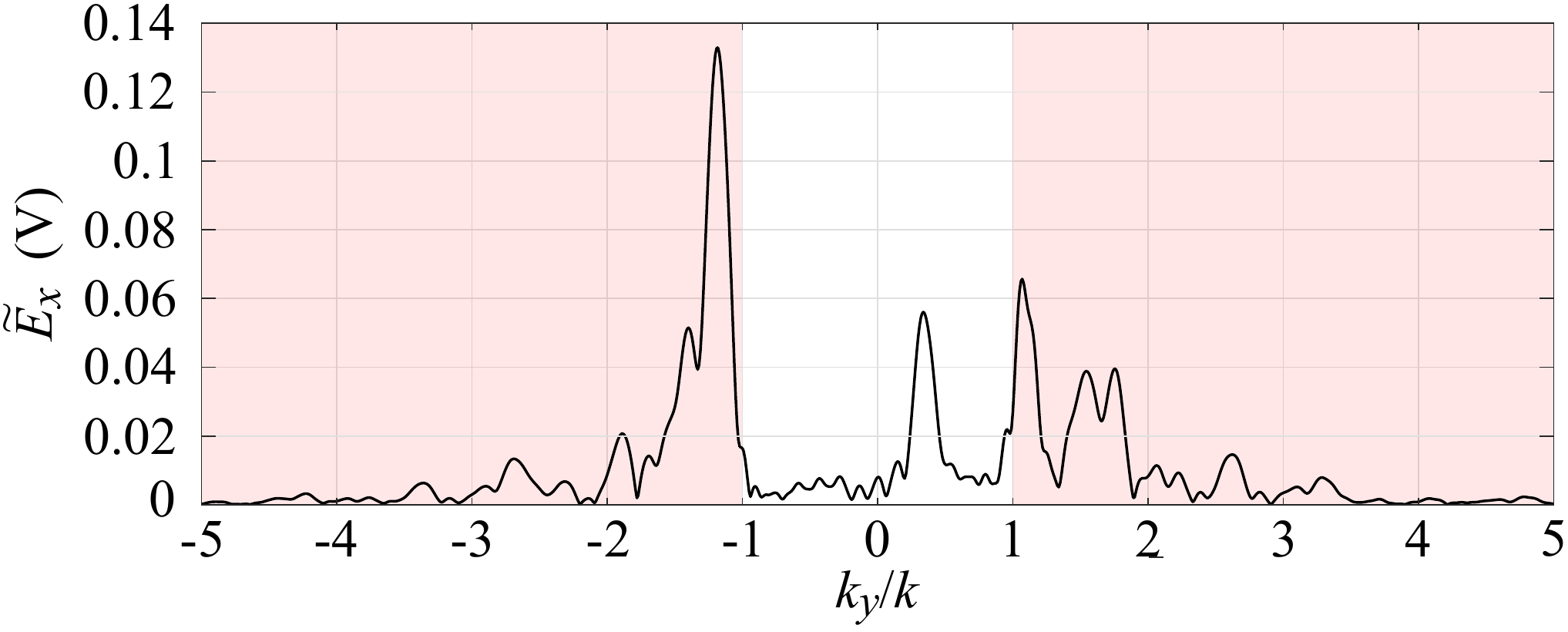}
\caption{Electric field spectrum for the MIMO impedance MTS antenna when its input located at $3\lambda$ is excited, evaluated at a plane $0.1\lambda$ above the meta-wires. The invisible region is shaded in red.}
\label{fig:MIMO_spectrum}
\end{figure}

The aperture electric field spectrum corresponding to the second output beam ($20^\circ$) is plotted in Fig.~\ref{fig:MIMO_spectrum}. Due to the symmetry of the device, the spectrum for the other output is simply a reflection of the plotted curve about $k_y=0$. As with the example examined in Sec.~\ref{sec:numerical_results:uniform_aperture}, we observe large Fourier components corresponding to tailored auxiliary surface waves inside the shaded invisible region. There is a high peak on the negative $k_y$ side of the spectrum, indicating strong SWs travelling towards the negative $y$-direction. This agrees with intuition, since the main role served by the SWs in this mode of operation is to deliver the source power from the right side to the left side of the aperture.

%%%%%%%%%%%%%%%%%%%%%%%%%%%%%%%%%%%%%%%%%%%%%%%%%%%%%%
%%%%%%%%%%%%%%%%%%%%%%%%%%%%%%%%%%%%%%%%%%%%%%%%%%%%%%
\section{Numerical Results with Realistic Meta-wires}
\label{sec:results:realistic}
In this section, we examine the physical realization of impedance MTSs featuring embedded sources and investigate their performance in terms of power efficiency and bandwidth. In order to converge to solutions that manifest higher bandwidth and milder sensitivity with respect to the wire impedances, we utilize all the constraints described in Sec.~\ref{sec:optimization:constraints}. After the optimization has been performed, the impedance loadings of each meta-wire are implemented with printed capacitors that are characterized in an aperiodic simulation, as detailed in App.~\ref{append:UnitCell}. These steps complete the end-to-end design of the embedded-source-fed impedance MTSs. Two design examples are presented, namely, an impedance MTS realizing a Chebyshev pattern with $-20 \ \mathrm{dB}$ side lobe level and impedance MTSs which radiate two distinct beams in two different directions.

\subsection{Chebyshev array pattern}
\label{sec:ChebyshevPattern}
In this design example, we realize a Chebyshev pattern with a constant side lobe level of $-20 \ \mathrm{dB}$ using a metasurface of width $W=7 \lambda$. The metasurface consists of $42$ wires placed on top of a grounded dielectric substrate with height $h=3.04 \mathrm{mm}$ and dielectric constant $\epsilon_r=3$, while the source is located at $(y_o,z_o)=(0,h/2)$. The geometrical parameters have been chosen so that two Rogers RO3003 substrates of $1.52 \ \mathrm{mm}$ thickness bonded together could realize the device. To determine the required far-field radiation, $14$ omni-directional virtual sources are assumed with their currents calculated based on the Chebyshev design method.

\begin{figure}[t]
\centering
{\includegraphics[width=0.98\columnwidth, trim=0 0 0 0 0, clip=true]{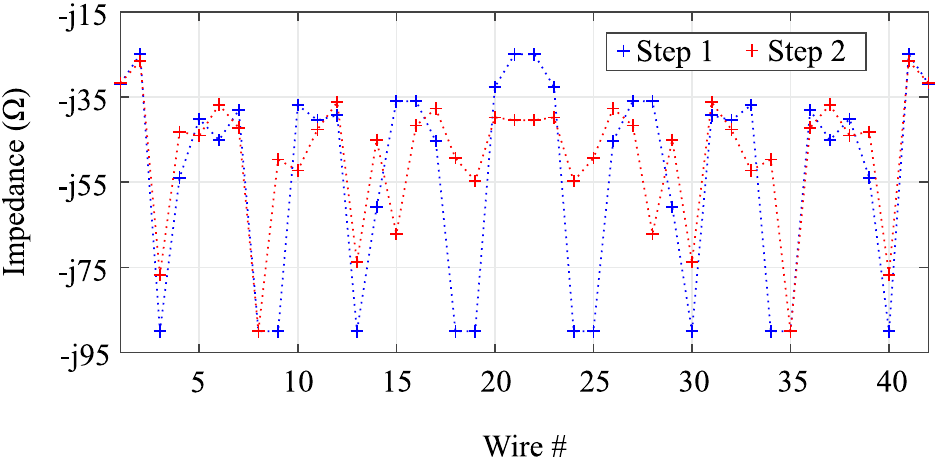}}
\caption{Optimized impedances of the meta-wires. The solution from Step 1 does not adhere to any nonlinear constraints, while the solutions at Step 2 is constrained in terms of Q-factor and sensitivity of the far-field.}
\label{fig:ChebyshevLoadings}
\end{figure}

The optimization is performed in two steps. In the first step, the impedances are constrained only by their passivity ($\mathrm{Re}\{Z_n\}=0$) and range ($[-j90,-j25]$). The converged solution is then used as the starting point of a second step that includes constraints on the Q-factor and sensitivity based on $Q_\mathrm{max}=40$ and $T_\mathrm{max}=40$ in \eqref{eqn:Qconstraint} and \eqref{eqn:SensitivityConstraint}, respectively. The converged values for the wire impedances of both steps are shown in Fig.~\ref{fig:ChebyshevLoadings}. These impedance values are then related to the width of the printed loading capacitors based on the design curve in Fig.~\ref{fig:WireCharacterization}, and simulations are performed in HFSS. The wire traces and ground plane are modelled with $18 \mathrm{um}$ copper and dielectric losses are introduced to the substrate through a loss tangent of $\mathrm{tan} (\delta)=0.001$.

\begin{figure}[t]
\centering
{\includegraphics[width=0.98\columnwidth, trim=0 0 0 0 0, clip=true]{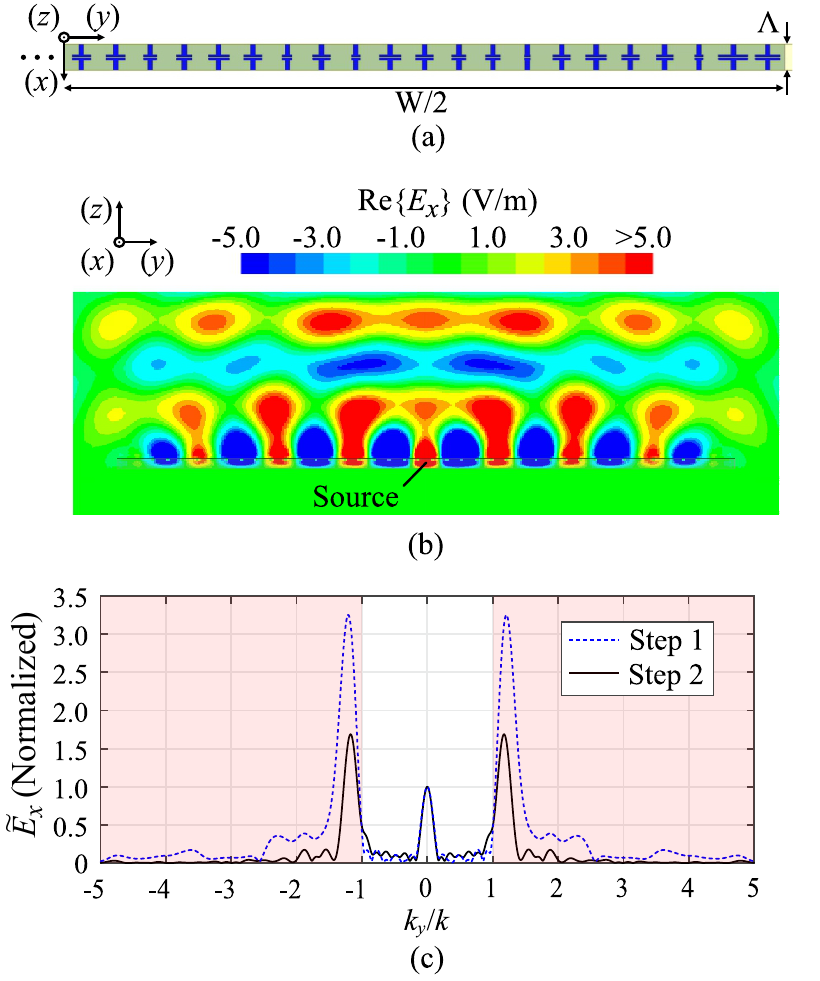}}
\caption{(a) Right half of one slice of the optimized Chebyshev beam-forming impedance MTS implemented using printed metallic wires, (b) Simulated near-field profile $\mathrm{Re}\{E_x\}$, and (c) Spectrum of the total electric field at a plane $0.1\lambda$ above the meta-wires (invisible region is shaded in red).}
\label{fig:ChebyshevFieldProfile}
\end{figure}

One half of a slice of the final optimized device implemented using realistic printed meta-wires is shown in Fig.~\ref{fig:ChebyshevFieldProfile}(a). Due to symmetry, the other half of the device is not shown. Its simulated near-field profile $\mathrm{Re}\{E_x\}$ is plotted in Fig.~\ref{fig:ChebyshevFieldProfile}(b). As seen, auxiliary surface waves are developed in the near-field of the metasurface, supported by the capacitive sheet above the grounded substrate. These surface waves symmetrically carry power towards the edges of the MTS antenna, so that an amplitude-taper required by a virtual Chebyshev array is obtained. 

From the HFSS simulation results, we also extract the spectrum of the electric field in a plane $0.1\lambda$ above the wires. We repeat this for the converged solutions of both optimization steps, and plot the resultant spectra normalized with respect to $\tilde{E}_x (k_y=0)$ in Fig.~{\ref{fig:ChebyshevFieldProfile}}(c). Although both solutions produce significant evanescent waves, it is clear that the additional constraints of the second optimization step succeeded in reducing the near-field reactive energy, which leads to a better radiation efficiency.

\begin{figure}[t]
\centering
{\includegraphics[width=0.98\columnwidth, trim=0 0 0 0 0, clip=true]{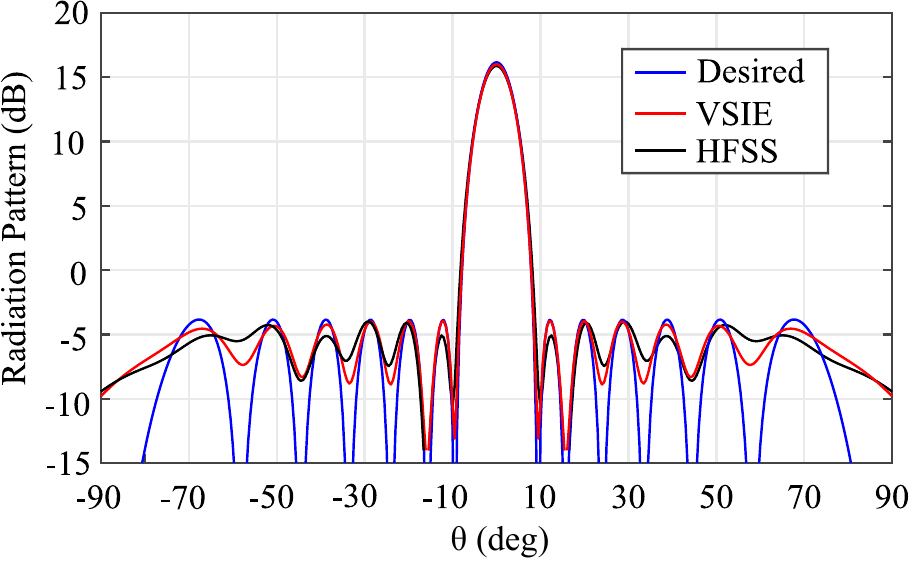}}
\caption{Radiation pattern of the Chebyshev antenna array at $10 \ \mathrm{GHz}$. VSIE (red curve) refers to the radiation pattern obtained from the optimized abstract wire loadings and HFSS (black curve) refers to the result from realistic simulations using loaded wires and including all power losses. Both match well with the targeted Chebyshev pattern (blue curve).}
\label{fig:ChebyshevRadiation}
\end{figure}

To assess the fidelity of the formed beam, we plot the simulated radiation pattern in Fig.~\ref{fig:ChebyshevRadiation}, along with the target pattern and the pattern predicted by the VSIE framework. As observed in Fig.~\ref{fig:ChebyshevRadiation}, the three patterns match each other very well. The deviation in simulated directivity compared to that of the targeted Chebyshev pattern is less than $ 0.3 \ \mathrm{dB}$. The realized the side lobe level is at $-19.9 \ \mathrm{dB}$, which is very close to the specified SLL of $-20$~dB. 

Notably, the excellent pattern matching was realized without any additional geometrical tuning of the printed capacitors. This is because the design curve of Fig.~{\ref{fig:WireCharacterization}} characterizes the intrinsic property of a single isolated meta-wire~\cite{Budhu:TAP2021}. The mutual coupling between meta-wires is rigorously and efficiently accounted by the VSIE framework. In other words, once established, the same design curve can be reused regardless of relative placement of the wires, as long as other geometrical and electromagnetic parameters remain unchanged. This feature of the VSIE-based design scheme means that it can be much more efficient and accurate than conventional methods which rely on inaccurate periodic surrogate models to characterize the unit-cells.

\begin{figure}[b]
\centering
{\includegraphics[width=0.98\columnwidth, trim=0 0 0 0 0, clip=true]{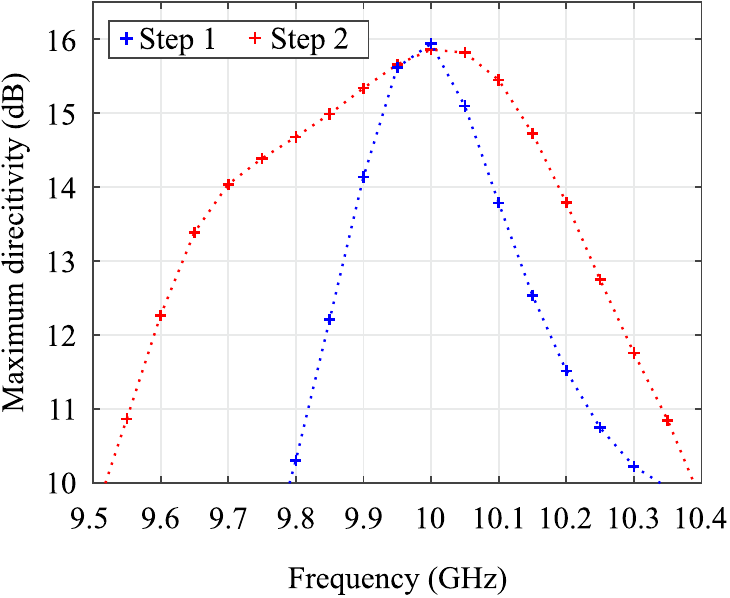}}
\caption{Frequency variation of the maximum directivity obtained from realistic wire simulations in HFSS. The final optimized solution (red curve) exhibits a wider bandwidth compared to the initial solution (blue curve) due to the additional constraints.}
\label{fig:Bandwidth}
\end{figure}

The realistic design allows for the investigation of the losses and the bandwidth. Specifically, losses are estimated to be around $7.4 \%$ ($3.3 \%$ ohmic losses in the wires and the ground plane and $4.1 \%$ in the dielectric). This constitutes a promising result, as multi-layer transmissive metasurfaces using auxiliary surface waves for similar functionality exhibit significantly higher losses \cite{Ataloglou:AWPL2021}. In addition, the frequency variation of directivity is plotted in Fig.~\ref{fig:Bandwidth}, showing a $3$-dB bandwidth of slightly over $6\%$ for the final optimized solution. The high power efficiency and acceptable bandwidth (considering the presence of surface waves) are attributed to not only the simplicity of the structure consisting of wires etched on a thin dielectric substrate, but also the additional nonlinear constraints that minimize the amplitude of the developed surface waves in the vicinity of the metasurface antenna. As a comparison, if we had concluded our optimization after the first step, the losses with a realistic structure would be $61.5\%$ and the $3$-dB bandwidth would be $2.6 \%$, as shown in Fig.~\ref{fig:Bandwidth}. This inspires confidence that even more broadband solutions can be obtained if tighter constraints are put in \eqref{eqn:Qconstraint} and \eqref{eqn:SensitivityConstraint}. However, this may require either slightly degrading the accuracy of the realized radiation pattern at the nominal frequency, or loosening some constraints (e.g. tolerating a higher SLL).

\subsection{Single-input Multi-output Metasurfaces}
\label{sec:numerical_results:multi_beam}
In this example, we examine metasurfaces, fed by a single embedded source, that realize two beams in two predefined directions. As in the previous example, the frequency is set to $10 \ \mathrm{GHz}$ and the metasurface consists of $42$ wires extending along an aperture size of $W=7 \lambda$. The substrate has a height of $h=3.04 \ \mathrm{mm}$ and dielectric constant $e_r=3$, while the source is placed at $(y_0,z_0)=(0,h/2)$. The desired radiation pattern is obtained by superimposing two uniform phased current sheets following ({\ref{eqn:uniform_aperture_radiation}}):
\begin{equation}
    \bar{J}_{des}[n]=A_1J_oe^{-jky_{gn}\sin\theta_{1}}+A_2J_oe^{-jky_{gn}\sin\theta_{2}}e^{j\xi}.
\end{equation}
Here, $A_1, A_2$ dictate the directivity of the two beams, $\theta_1, \theta_2$ are the two output angles, and $\xi$ is a constant phase that slightly modifies the side lobes of the total pattern by affecting the interference of the two beams. For our examples, we use $A_1=A_2=1$ resulting in desired patterns with two equally directive beams. It is emphasized that both beams originate from a single source, unlike the MIMO example presented in Sec.~\ref{sec:numerical_results:MIMO}. In a sense, the present MTSs can be perceived as single-input multiple-output (SIMO) systems, as two spatially separated receivers could receive the same transmitted signal. With that in mind, we present two design examples with the specifications shown in Table~\ref{table:SIMO_spec}.

\begin{table}[h]
\begin{center}
		\caption{Design specifications for two different SIMO MTSs.}
		\label{table:SIMO_spec}
 		\begin{tabular}{cccc}
 		\hline
 		\hline
 		& $\theta_1$& $\theta_2$ & $\xi$\\
 		\hline
		Design A & $-45^\circ$ & $-10^\circ$ & $\pi/2$\\
		\hline
		Design B & $-45^\circ$ & $22.5^\circ$ & $0$\\
		\hline
		\hline
		\end{tabular}
\end{center}
\end{table}

\begin{figure}[t]
\centering
{\includegraphics[width=0.98\columnwidth, trim=0 0 0 0 0, clip=true]{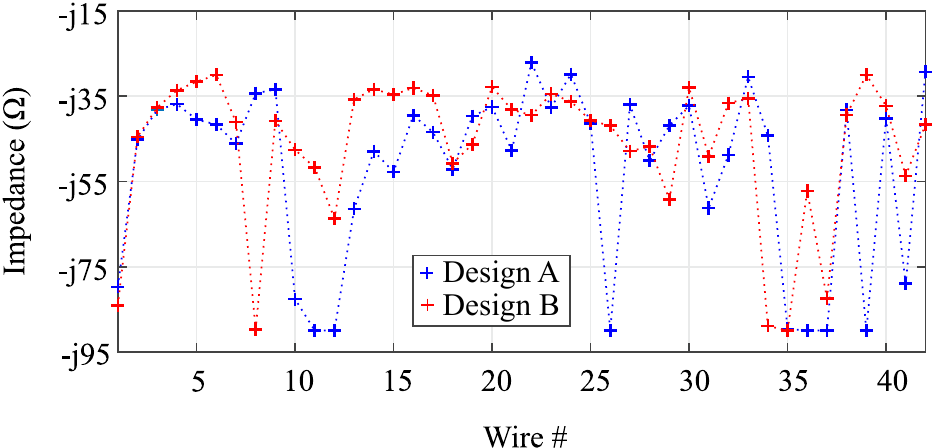}}
\caption{Optimized meta-wire impedances for the two SIMO MTS designs.}
\label{fig:Multibeam_Impedances}
\end{figure}

The loading impedances of the wires for each case are determined based on the same two-step approach. First, a gradient-descent optimization is ran aiming at passive loadings in the $[-j90, -j25]$ range. The converged solutions are then used as a starting point in the second optimization step with the additional constraints $Q_\mathrm{max}=50$ and $T_\mathrm{max}=65$. These limits were selected to obtain the best compromise between pattern fidelity and device bandwidth as well as power efficiency. The final values for the meta-wire impedances in each of the two cases are depicted in Fig.~\ref{fig:Multibeam_Impedances}. These impedance values are mapped to printed capacitor sizes based on Fig.~\ref{fig:WireCharacterization} in Appendix~\ref{append:UnitCell}.

\begin{figure}[b]
\centering
{\includegraphics[width=0.98\columnwidth, trim=0 0 0 0 0, clip=true]{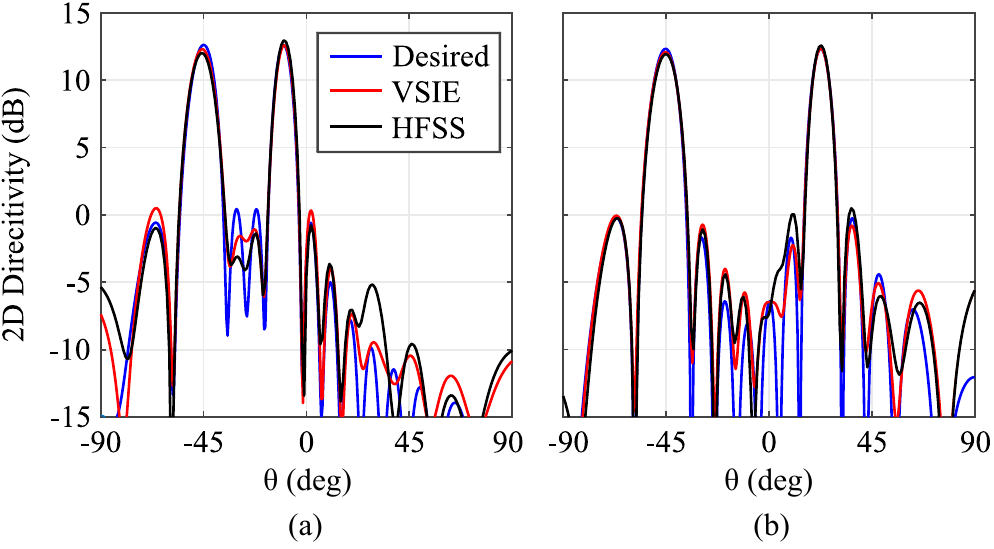}}
\caption{Radiation patterns for (a) Design A and (b) Design B. The full-wave simulated results (black) show good agreement with the optimized VSIE (red) and the desired (blue) radiation patterns in terms of the beam angles and peak directivities.
}
\label{fig:Multibeam_Radiation}
\end{figure}

Simulations using realistic wires and losses are performed in HFSS. \iffalse The near-field patterns are given in Fig.~\ref{fig:Multibeam_NF}, where the developed surface waves supported by the loaded wires are clearly visible. Moreover, above the MTS, an interference pattern is formed by the two radiated beams in the different directions. \fi
To assess the beam-forming accuracy, the 
realized far-field patterns for both designs are plotted in Fig.~\ref{fig:Multibeam_Radiation}. The patterns predicted by the VSIE framework are also included for comparison. Evidently, both designs produced output beams at their designated angles. For all four output beams, the full-wave simulated (maximum) directivity values differ from the desired values by no more than 0.6~dB.  Furthermore, both designs have side lobe levels less than -11.5~dB, which are very close to the desired side lobes when converted to linear scale.

The nonlinear constraint on the Q-factor helped us obtain power efficiencies of $92.3\%$ for Design~A and $91.0\%$ for Design~B. The power efficiency is limited by dielectric losses ($4.2 \%$ for A and $4.5\%$ for B) and ohmic losses in copper ($3.5 \%$ for A and $4.5 \%$ for B). For each design, we examine its worst case 3-dB bandwidth, i.e. the bandwidth of its more narrow band beam. For Design A and Design B, these are evaluated to be $4.5 \%$ and $3.9 \%$ respectively.

%%%%%%%%%%%%%%%%%%%%%%%%%%%%%%%%%%%%%%%%%%%%%%%%%%%%%%
%%%%%%%%%%%%%%%%%%%%%%%%%%%%%%%%%%%%%%%%%%%%%%%%%%%%%%
\section{Conclusion}
\label{sec:conclusion}
We presented the design of a complete low-profile beam-forming platform consisting of a single-layered ground-plane-backed impedance MTS fed by embedded sources. Through the use of an efficient optimization-based design strategy derived from volume-surface integral equations, several devices with extreme beam-forming capabilities were synthesized. By harnessing tailored auxiliary surface waves, they were able to realize optimal aperture illumination regardless of their physical size or the source type, while demonstrating useful functionalities such as wide-angle or MIMO beam-forming. To bring the proposed devices one step closer to physical realization, we derived several feasibility-related constraints which serve to enhance their bandwidths and robustness against fabrication errors, while reducing ohmic and dielectric losses. With the help of a simple mapping procedure between theoretical MTS models and realistic PCB-compatible devices, we constructed several designs capable of arbitrary beam-shaping or SIMO beam-forming using simple capacitively loaded printed copper traces. Full-wave simulations corroborates the exceptional theoretically predicted beam-forming capabilities, while confirming the effectiveness of the proposed feasibility constraints.

\begin{appendices}
\section{Characterization of wire impedances}
\label{append:UnitCell}
\begin{figure}[t]
\centering
{\includegraphics[width=0.98\columnwidth, trim=0 0 0 0 0, clip=true]{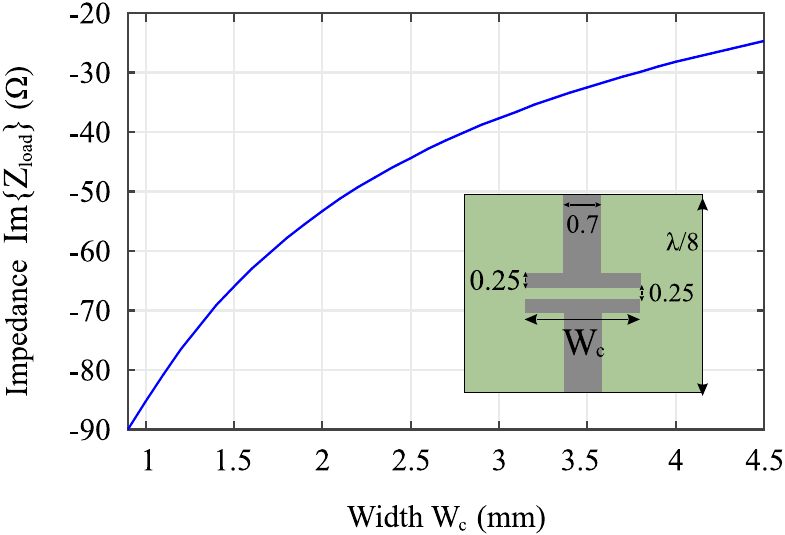}}
\caption{Extracted wire impedance values for capacitively-loaded wires as a function of the width $W_c$ (inset: Geometry of printed loading. Absolute dimensions are in $\mathrm{mm}$).}
\label{fig:WireCharacterization}
\end{figure}

The meta-wires comprising the unit-cells of our MTSs are periodically loaded in the longitudinal direction (every $\lambda/8$) with printed capacitors, as seen in the inset of Fig.~\ref{fig:WireCharacterization}. By varying the width $W_c$ of the capacitor, different effective wire impedances are acquired. 

To characterize the impedance of a particular wire design, we employ a modified version of a method previously used to characterize meta-wires suspended in air~\cite{Budhu:TAP2021}. We first consider a hypothetical device which has the same grounded dielectric substrate and source as the actual MTS design, but with only a single wire printed at $(y,z)=(0,h)$. Through a full-wave simulation in HFSS, the total field radiated by this single-wire device is evaluated along a near-field observation segment. We then consider the VSIE model for this device, in which the printed wire is replaced by a homogeneous sheet with unknown surface impedance $Z_w$. With the method presented in Sec.~{\ref{sec:model}}, it is very easy to evaluate the theoretically expected electric field along the near-field observation segment for various values of $Z_w$. The value of $Z_w$ that gives the best match between VSIE and HFSS results is designated as the effective impedance of the meta-wire design under consideration. Importantly, this method characterizes the wire in the presence of the grounded dielectric substrate, which has a non-negligible impact on the effective capacitance of the printed capacitor.

Aiming at the realistic implementation of the design example in Sec.~\ref{sec:ChebyshevPattern}, we characterize the loaded wires above a dielectric of height $h=3.04 \mathrm{mm}$ and $\epsilon_r=3$. The ground plane is also present and the whole structure has a finite width of $7 \lambda$. The extracted effective impedances for various values of capacitor width $W_c$ is plotted in Fig.~\ref{fig:WireCharacterization}. As expected, widening the printed loading increases the capacitance, or equivalently, decreases the effective reactance. For applications that require more negative reactances, it is sufficient to increase the gap of the capacitor. More positive values can be obtained with other types of wires such as those loaded with meandering inductors~\cite{Budhu:TAP2021}. However, for our design examples, the obtained range of $[-j90,-j25] \ \Omega$ was adequate. In addition, no significant change was observed when adding or removing copper losses, indicating that the wires operate away from their resonance. 

\end{appendices}

\ifCLASSOPTIONcaptionsoff
  \newpage
\fi

\bibliography{references} 
\bibliographystyle{ieeetr}
\end{document}